\documentclass[preprint,12pt]{elsarticle}

\usepackage{amssymb}
\journal{Information Fusion}

\makeatletter
\newtoks\@eadauthorshort
\def\@author#1#2{\g@addto@macro\elsauthors{\normalsize%
		\def\baselinestretch{1}%
		\upshape\authorsep#1\unskip\textsuperscript{%
			\ifx\@fnmark\@empty\else\unskip\sep\@fnmark\let\sep=,\fi
			\ifx\@corref\@empty\else\unskip\sep\@corref\let\sep=,\fi
		}%
		\def\authorsep{\unskip,\space}%
		\global\let\@fnmark\@empty
		\global\let\sep\@empty}%
	\@eadauthor={#1}
	\@eadauthorshort={#2}
}
\def\@@author[#1]#2#3{\g@addto@macro\elsauthors{%
		\def\baselinestretch{1}%
		\authorsep#2\unskip\textsuperscript{%#1%
			\@for\@@affmark:=#1\do{%
				\edef\affnum{\@ifundefined{X@\@@affmark}{1}{\elsRef{\@@affmark}}}%
				\unskip\sep\affnum\let\sep=,}%
			\ifx\@fnmark\@empty\else\unskip\sep\@fnmark\let\sep=,\fi
			\ifx\@corref\@empty\else\unskip\sep\@corref\let\sep=,\fi
		}%
		\def\authorsep{\unskip,\space}%
		\global\let\sep\@empty\global\let\@corref\@empty
		\global\let\@fnmark\@empty}%
	\@eadauthor={#2}%
	\@eadauthorshort={#3}%
}
\gdef\@ead#1{\bgroup\def\_{\string\underscorechar\space}%
	\def\{{\string\lbracechar\space}%
	\def~{\hashchar\space}%
	\def\}{\string\rbracechar\space}%
	\edef\tmpA{\the\@eadauthor}
	\edef\tmpB{\the\@eadauthorshort}
	\immediate\write\@auxout{\string\emailauthor
		{#1}{\expandafter\strip@prefix\meaning\tmpA}{\expandafter\strip@prefix\meaning\tmpB}}%
	\egroup
}
\gdef\emailauthor#1#2#3{\stepcounter{ead}%
	\g@addto@macro\@elseads{\raggedright%
		\let\corref\@gobble
		\eadsep\texttt{#1} (\ifemailshortauthor #3\else#2\fi)\def\eadsep{\unskip,\space}}%
}
\newif\ifemailshortauthor
\makeatother

\emailshortauthortrue

\usepackage{subfigure}
\usepackage{amsmath}
\usepackage{multirow}
\usepackage{multicol}
\usepackage{makecell}
\usepackage{color,xcolor}
\usepackage{array, caption, threeparttable}
\usepackage[labelfont=bf,labelsep=none]{caption}
\begin{document}

\begin{frontmatter}



\title{Multispectral Pan-sharpening via Dual-Channel Convolutional Network with Convolutional LSTM Based Hierarchical Spatial-Spectral Feature Fusion}


\author[1]{Dong Wang}{D. Wang}
\ead{dongwang@mail.nwpu.edu.cn}
\author[2]{Yunpeng Bai}{Y. Bai}
\ead{yunpengb@student.unimelb.edu.au}
\author[1]{Ying Li\corref{cor1}}{Y. Li}
\ead{lybyp@nwpu.edu.com}

\cortext[cor1]{Corresponding author}

\address[1]{School of Computer Science, National Engineering Laboratory for Integrated Aero-Space-Ground-Ocean Big Data Application Technology, Shaanxi Provincial Key Laboratory of Speech \& Image Information Processing, Northwestern Polytechnical University, Xi’an, China}
\address[2]{School of Computing and Information Systems, The University of Melbourne, Victoria 3010, Australia.}

\begin{abstract}
	Multispectral pan-sharpening aims at producing a high resolution (HR) multispectral (MS) image in both spatial and spectral domains by fusing a panchromatic (PAN) image and a corresponding MS image. In this paper, we propose a novel dual-channel network (DCNet) framework for MS pan-sharpening. In our DCNet, the dual-channel backbone involves a spatial channel to capture spatial information with a 2D CNN, and a spectral channel to extract spectral information with a 3D CNN. This heterogeneous 2D/3D CNN architecture can minimize causing spectral information distortion, which typically happens in conventional 2D CNN models. In order to fully integrate the spatial and spectral features captured from different levels, we introduce a multi-level fusion strategy. Specifically, a spatial-spectral CLSTM (S$^2$-CLSTM) module is proposed for fusing the hierarchical spatial and spectral features, which can effectively capture correlations among multi-level features. The S$^2$-CLSTM module attaches two fusion ways: the intra-level fusion via bi-directional lateral connections and inter-level fusion via the cell state in the S$^2$-CLSTM.  Finally, the ideal HR-MS image is recovered by a reconstruction module. Extensive experiments have been conducted at both simulated lower scale and the original scale of real-world datasets. Compared with the state-of-the-art methods, the proposed DCNet achieves superior or competitive performance.
\end{abstract}

\begin{graphicalabstract}
\includegraphics[width=\linewidth]{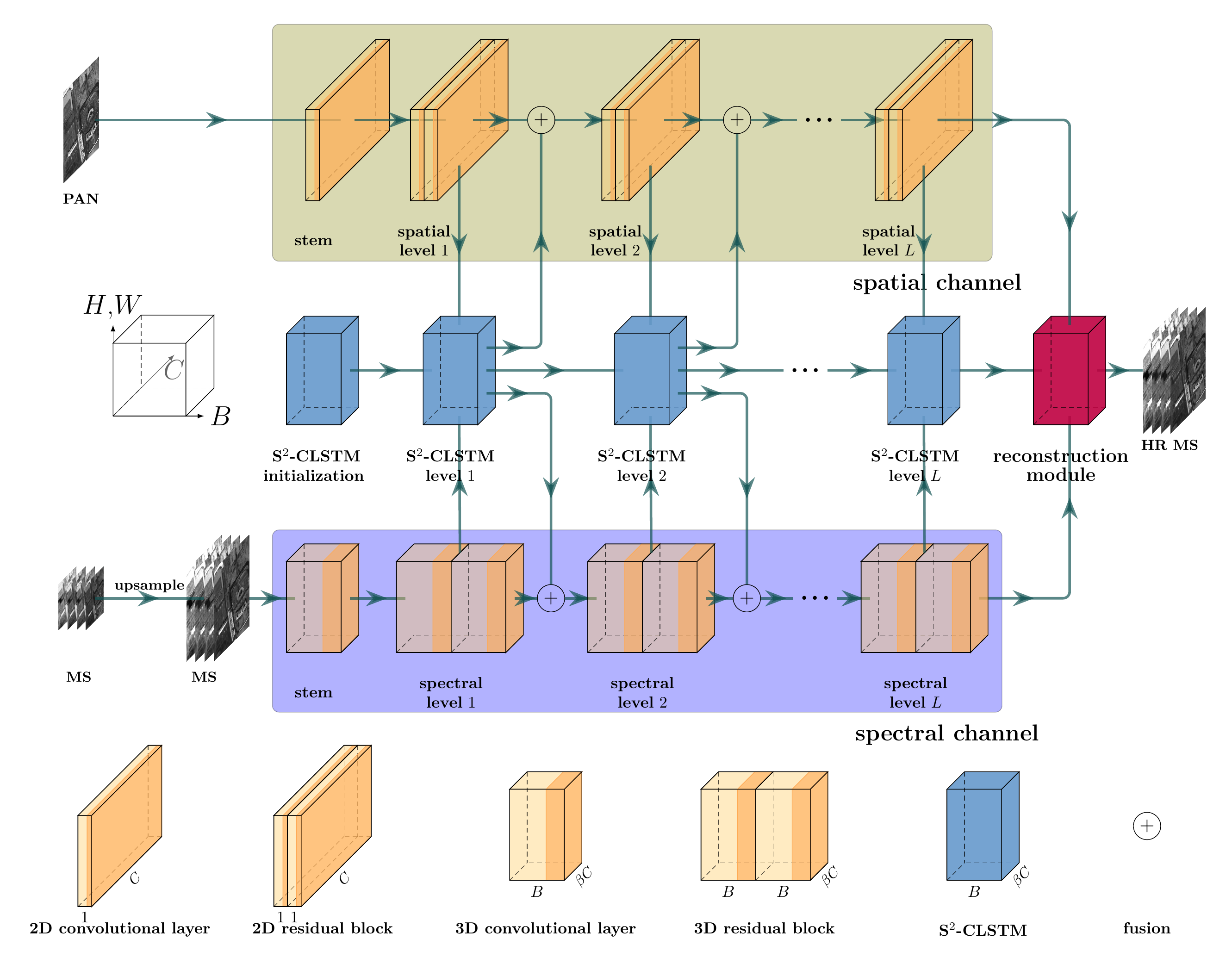}
\end{graphicalabstract}

\begin{highlights}
\item A heterogeneous dual-channel backbone is proposed for feature extraction
\item The proposed DCNet adopts a multi-level fusion strategy
\item An S$^2$-CLSTM module is proposed to fuse hierarchical spatial and spectral features
\item Extensive experiments have been conducted at the lower and original scales

\end{highlights}

\begin{keyword}


Dual-Channel Network(DCNet) \sep Convolutional Long Short Term Memory (CLSTM)
\sep Hierarchical Fusion \sep Multispectral Pan-sharpening

\end{keyword}

\end{frontmatter}


\section{Introduction}
\label{Introduction}

High resolution (HR) multispectral (MS) images in both spectral and spatial domains are desirable for many practical applications, such as environmental monitoring \cite{bullock2020monitoring}, object detection \cite{gong2019context}, and classification \cite{fang2018semi,li2017spectral,fang2019hyperspectral}. However, due to the hardware limitations, it is hard to provide such ideal images, and only panchromatic (PAN) images with high spatial resolution and low spatial resolution MS images can be captured by sensors, e.g., IKONOS, GaoFen-2, and WorldView-2. Multispectral pan-sharpening refers to the technology of obtaining the HR-MS image by fusing the PAN image and the corresponding MS image \cite{liu2018deep}.

During the last decades, many and various MS pan-sharpening methods have been proposed, e.g., component substitution (CS) based methods \cite{garzelli2007optimal,aiazzi2007improving}, multi-resolution analysis (MRA) based methods \cite{khan2008indusion,ranchin2003image}, model-based methods \cite{palsson2019model,fu2019variational}, etc. CS-based methods assume that spatial information and spectral information are separable. Firstly, they up-sample MS images to the same spatial resolution with PAN images, and then these MS and PAN images are projected into another feature domain. Finally, the LR component of MS images is substituted by the counterpart of PAN images \cite{thomas2008synthesis}. The underlying assumption of MRA methods is that high-frequency details lacked in MS images can be supplemented by PAN images. So they extract high-frequency information of PAN images and inject it into MS images while retaining all original spectral information of MS images. Generally speaking, model-based methods can obtain fused images of relatively high quality. However, they are time-consuming for inferring the desired HR-MS images because optimization algorithms usually are computationally intensive.

\begin{figure}[t]
	\setlength{\abovecaptionskip}{-0.5 cm}
	\begin{center}
		\subfigure [Ground truth]{
			\label{fig:spatial_and_spectral_distortions--ground_truth}
			\includegraphics[width=0.3\linewidth]{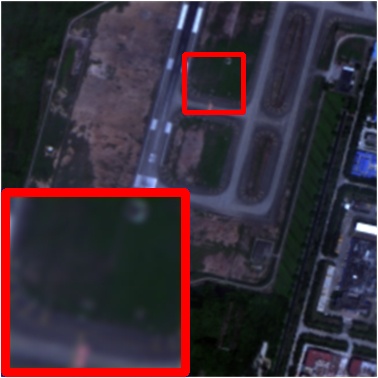}
		}
		\subfigure [Spatial distortion]{
			\label{fig:spatial_and_spectral_distortions--spatial_distortion}
			\includegraphics[width=0.3\linewidth]{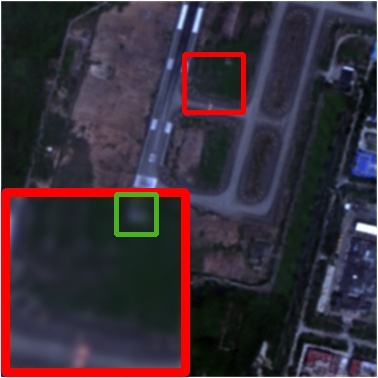}
		}
		\subfigure [Spectral distortion]{
			\label{fig:spatial_and_spectral_distortions--spectral_distortion}
			\includegraphics[width=0.3\linewidth]{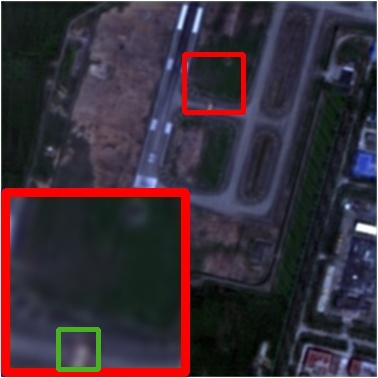}
		}
	\end{center}
	\caption{An example of spatial and spectral distortions. The left image is the ground truth. The green bounding boxes indicate these distortions.}
	\label{fig:spatial_and_spectral_distortions}
\end{figure}

Recently, a large number of convolutional neural network (CNN) based methods have been proposed for pan-sharpening and have shown great potential. The high nonlinearity of deep CNNs facilitates modeling the fusion of PAN and MS images. Although these methods have achieved promising performance, there is still some space for reducing the spatial and spectral distortions. For example, the L-shaped object in the green bounding box of Fig. \ref{fig:spatial_and_spectral_distortions--spatial_distortion} is not the same with the lunar shape of ground truth in Fig. \ref{fig:spatial_and_spectral_distortions--ground_truth}, which is referred to spatial distortion. The spectral distortion can be inferred by the color difference between the ground truth and the fused image, e.g., the color of the rectangle object in Fig. \ref{fig:spatial_and_spectral_distortions--spectral_distortion} is whiter than the pink ground truth in Fig. \ref{fig:spatial_and_spectral_distortions--ground_truth}.

Spatial and spectral distortions in the pan-sharpened MS image comes from many reasons. One of them is that most methods employ 2D CNNs as a spectral feature extractor for 3D MS images. 2D CNN is an effective spatial feature extractor for 2D PAN images. When it comes to 3D MS images, it will mix different spectrums of MS images and will make it challenging to recover each band of HR-MS images from 2D feature maps. Besides, most existing methods only fuse spatial and spectral information at an early stage or a late stage. The hierarchical CNN features, outstanding powerful features, need to be exploited for spatial and spectral information fusion. A better fusion of these hierarchical features can significantly contribute to reducing the spatial and spectral distortions.

To this end, we proposed a dual-channel network (DCNet) that includes three main components: the dual-channel backbone for spatial and spectral feature extraction, a novel spatial-spectral CLSTM (S$^2$-CLSTM) for hierarchical fusion, and the reconstruction module for HR-MS image synthesis. Specifically, our model first learns a dual-channel backbone to capture spatial and spectral information from 2D PAN images and 3D MS images, respectively.  Then, the extracted hierarchical spatial and spectral features are fused by the S$^2$-CLSTM module which can carry out both the intra-level and the inter-level fusion. Finally, our DCNet combines a reconstruction module to generate the ideal HR-MS images. The main contributions of this paper are summarized as follows:

\begin{itemize} 
	\item A heterogeneous dual-channel backbone is proposed for feature extraction in multispectral pan-sharpening. The spatial channel is designed to capture spatial details of 2D PAN image with a 2D CNN, while the spectral channel is responsible for obtaining spectral information in 3D MS images with a 3D CNN.
	
	\item Different from most existing CNN based pan-sharpening approaches that employ the single-level fusion strategy, i.e., early-fusion or late-fusion. The proposed DCNet adopts a hierarchical fusion strategy to integrate spatial and spectral features level by level.
	
	\item An S$^2$-CLSTM module is proposed to effectively capture correlations among hierarchical spatial and spectral features and fully integrate these features. In S$^2$-CLSTM, the intra-level and inter-level spatial and spectral feature fusion are carried out via bi-directional lateral connections and the cell state of S$^2$-CLSTM, respectively. To the best of our knowledge, CLSTM is employed in the field of pan-sharpening for the first time.
	
	\item Extensive experiments on three datasets, e.g., IKONOS, GaoFen-2, and WorldView-2, have been conducted at both the lower scale and the original scale. At the lower scale experiments, the proposed DCNet substantially outperforms state-of-the-art methods. Experiment results at the original scale also demonstrate that the DCNet achieves superior or competitive performance.
\end{itemize}

The remainder of this paper is organized as follows. Section \ref{Related Work} briefly introduces the background knowledge of the CLSTM and the existing CNN-based pan-sharpening methods. A detailed description of the proposed method is presented in Section \ref{Methodology}. The experimental results are presented and discussed in Section \ref{Experiments}. Finally, the conclusion of this paper is given in Section \ref{Conclusion}.

\section{Related Work}
\label{Related Work}

\subsection{CNN for multispectral pan-sharpening}
Recently, CNN has become very popular in the pan-sharpening society. Most of the CNN-based methods employ 2D CNNs for feature extraction from both PAN and MS images. Besides, spectral and spectral information are normally fused at an early stage or a late stage. Few methods have realized the potential of the hierarchical spatial and spectral features. These methods are described below.

Inspired by SRCNN \cite{dong2015image}, Masi et al. \cite{masi2016pansharpening} proposed PNN, which is the first model based on CNN. In the underlying architecture of PNN, the LR MS image is first upsampled and interpolated. The upsampled MS bands and the PAN band are then concatenated and are tailored by a three-layer 2D CNN, which means that the network fuse the PAN and MS images at the early stage. PNN works at HR from the beginning, and the output comprises four bands corresponding to HR-MS images. Scarpa et al. \cite{scarpa2018target} extended residual learning to the PNN and obtained a significant performance gain over PNN. Different from PNN, the proposed PNN+ has a skip connection from input to the output of the network. Besides, a target-adaptive tuning phase is introduced to solve the problem of insufficient data and allows users to apply the proposed architecture to their dataset. Wei et al. \cite{wei2017boosting} introduced a deep convolutional neural network with residual learning (DRPNN). The network takes the concatenation of PAN and MS images, which is the same with PNN. DRPNN \cite{wei2017boosting} is a very deep convolutional neural network to make full use of the high nonlinearity of deep learning models. All of these networks employ 2D CNNs for pan-sharpening, and they fuse MS and PAN images at the early stage.

Yang et al. \cite{yang2017pannet} designed a pan-sharpening network called PanNet that takes the high-pass components of the PAN images and MS images instead of original images. In PanNet, domain-specific knowledge is incorporated to preserve spectral and spatial information. For spectral preservation, they add up-sampled multispectral images to the network output, which directly propagates the spectral information to the reconstructed image. For spatial maintenance, the network is trained in the high-pass filtering domain rather than the image domain, the input is the concatenated high-pass components of the PAN and upsampled LR MS images. So the spatial and spectral information fusion are fused at an early stage in this method. Later, the multi-scale and multi-depth convolutional neural network (MSDCNN) is proposed by Yuan et al. \cite{yuan2018multiscale}. They also concatenated the PAN band and the MS bands together and fed it into the network, which is the same with PNN. But they employ a shallow and a deep multi-scale branch to model pan-sharpening.

Unlike the methods mentioned above, Liu et al. \cite{liu2020remote} proposed a two-stream fusion network (TFNet) that extracts CNN features from PAN and MS images, and then fuse them at the late stage. TFNet has three modules whose functions are feature extraction, feature fusion, and image reconstruction, respectively. TFNet firstly extracts spatial and spectral features by two 2D CNNs and then fuses the elements. All convolutional layers in TFNet are 2D, and they only fuse the features of two streams at the late stage. Zhang et al. \cite{zhang2019pan} presented a bi-directional pyramid network (BDPN) for pan-sharpening, which fuses the features of PAN and MS images at two stages. But they inject the spatial information of PAN images into MS images in the image domain instead of the feature domain, which prevents the BDPN from leveraging the hierarchical features. Shao et al. \cite{shao2018remote} proposed a remote sensing image fusion named RSIFNN that can adequately extract spectral and spatial features from source images. Although they studied the effect of different depths of each branch, they only fuse the spatial and spectral features at the late stage and fail to leverage the hierarchical features of PAN and MS branches.

\subsection{CLSTM}

The Long short-term memory (LSTM) has achieved great success for sequence modeling in various natural language processing tasks, e.g., language processing, speech recognition \cite{graves2013speech}, and visual question answering\cite{bai2020decomvqanet}. With the cell state and gates, LSTMs can remove or add to cell state and remember long term dependencies. However, LSTMs only take as input 1-D vectors and thus cannot be applied for 2D feature maps. Shi et al. \cite{xingjian2015convolutional} introduced 2D convolution operation to LSTM and proposed CLSTM, which can process 2-D feature maps and automatically capture temporal dependencies.

CLSTMs can also be used for 3D data processing. Song et al. \cite{song2018pyramid} proposed a fast video salient object detection model, based on pyramid dilated bidirectional ConvLSTM (PDB-ConvLSTM). In \cite{liu2016spatio}, Liu et al. proposed a powerful tree-structure based traversal method to model the 3D-skeleton and CLSTM to handle the noise and occlusions in 3D skeleton data. Jiang et al. \cite{jiang2017predicting} proposed an object-to-motion convolutional neural network (OM-CNN). In the model, a two-layer convolutional long short-term memory (2C-LSTM) network to predict video saliency.

\section{Methodology}
\label{Methodology}

In this section, we first present the proposed DCNet. Then we illustrate details of three main components of DCNet: the dual-channel backbone, S$^2$-CLSTM, and the reconstruction module. The objective function is introduced at last.

\subsection{Fusion framework}

The purpose of multispectral pan-sharpening is to get an HR-MS image by fusing a PAN image and a corresponding MS image, which has $B$ bands (e.g., $B=4$ for IKONOS and GaoFen-2 satellite, while $B=8$ for WorldView-2 satellite). In this paper, the observed PAN image is denoted as ${{X}_{P}}\in {{R}^{H\times W}}$, where $H$ and $W$ are the height and width of the PAN image, respectively. ${{X}_{M}}\in {{R}^{\frac{H}{4}\times \frac{W}{4}\times B}}$ represents the corresponding MS image, with 4 being a spatial reduction ratio. We denote the pan-sharpened HR-MS image as ${{Y}_{M}}\in {{R}^{H\times W\times B}}$. 

\begin{figure}[t]
 	\setlength{\abovecaptionskip}{-0.5 cm}
 	\begin{center}
 		\includegraphics[width=1\linewidth]{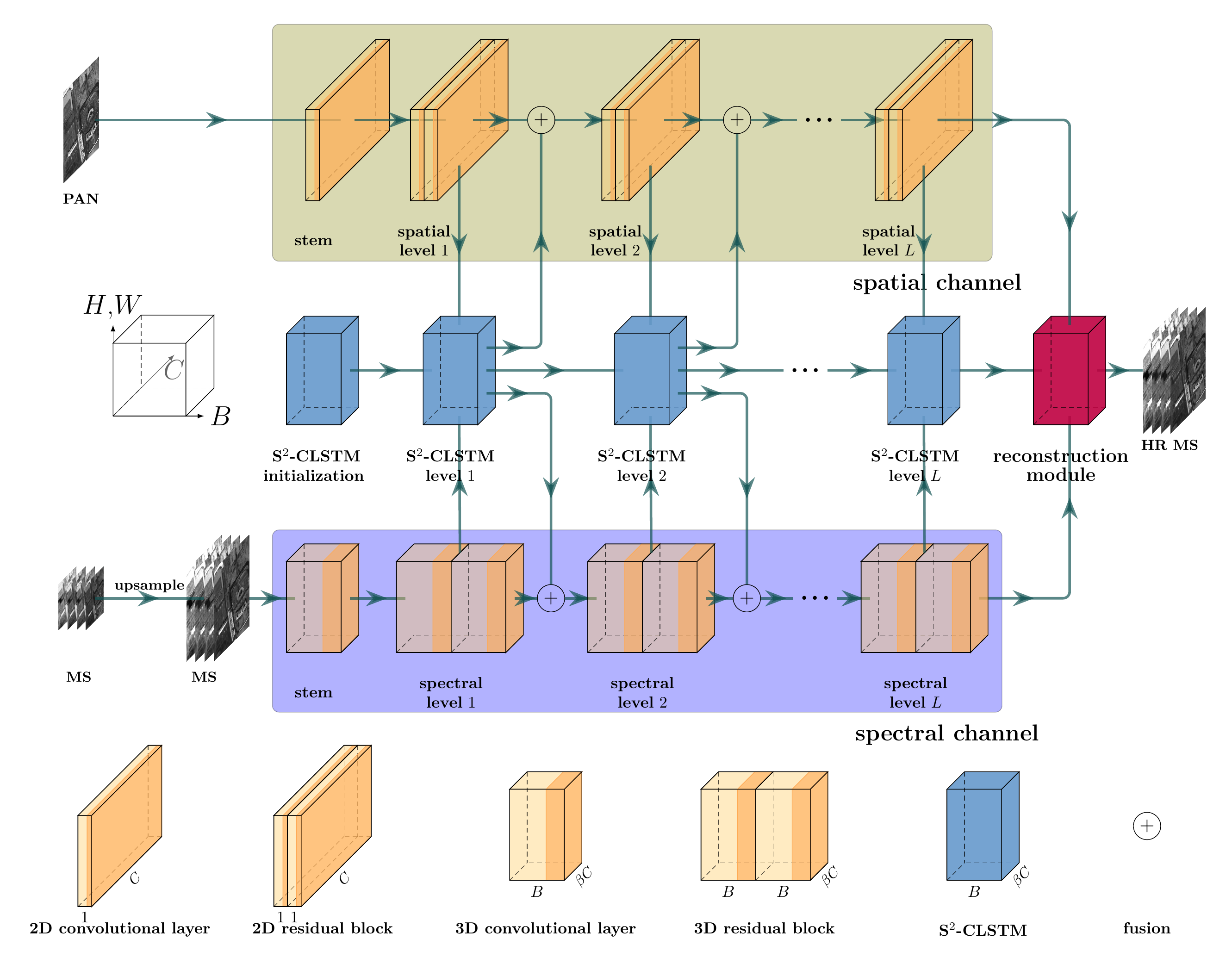}
 	\end{center}
 	\caption{The architecture of the proposed DCNet for multispectral pansharpning. $H$ and $W$ represent the height and width, respectively. $B$ indicates the number of bands. $C$ denotes the number of channels. $\beta$ is the filter number ratio between the spatial channel and the spectral channel.}
 	\label{fig:architecture}
 \end{figure}

A detailed illustration of the proposed DCNet can be found in Fig. \ref{fig:architecture} Specifically, a dual-channel backbone is utilized to obtain the hierarchical spatial and spectral features, which is tailed the PAN and MS images. Then, the S$^2$-CLSTM fuses spatial and spectral information and is located in the middle of the two channels in the figure. Finally, the reconstruction module takes the high-level features of the spatial channel, the spectral channel, and the S$^2$-CLSTM as input and synthesis the desired HR-MS image. The coordinate system indicating the dimension of width, height, bands, and channels is at the left of Fig. \ref{fig:architecture}.

\subsection{The dual-channel Backbone}
\label{TheDual-channelBackbone}
\subsubsection{Spatial channel}

The spatial channel contains a stem layer and stacked 2D residual blocks. The stem layer is comprised of a 2D convolutional layer and a parametric ReLU (PReLU). The first residual block is represented as $R_{2D}^{1}(\cdot )$, where 1 denotes the first residual block. The other residual blocks are represented as $R_{2D}^{l}(\cdot )$, so the spatial channel can be formulated as follows:
\begin{equation}F_{P}^{0}=PReLU(w_{P}^{0}*{{X}_{P}}+b_{P}^{0})\end{equation}
\begin{equation}F_{P,CLSTM}^{l}=R_{2D}^{l}(F_{P}^{l-1})\end{equation}
\begin{equation}F_{P}^{l}=F_{P,CLSTM}^{l}+F_{CLSTM,P}^{l}\end{equation}
\begin{equation}F_{P}^{L}=R_{2D}^{L}(F_{P}^{L-1})\end{equation}
where $F_{P}^{0}$ is the spatial feature extracted by the stem layer, $F_{CLSTM,P}^{l}$ comes from the level $l$ of S$^2$-CLSTM, and $l$ ranges from 1 to $L-1$, and $PReLU$ is a type of leaky ReLU. The 2D residual blocks can be formulated as
\begin{equation}R_{2D}^{l}(x)=x+PReLU(w_{P}^{l,1}*PReLU(w_{P}^{l,0}*x+b_{P}^{l,0})+b_{P}^{l,1})\end{equation}
where $x$ is the input of residual blocks, $w$ and $b$ are learnable parameters, and $l$ indexes the $l$-th residual block. Each residual block has two successive convolutional layers.

\subsubsection{Spectral channel}
The overall architecture of the spectral channel is the same as the spatial channel. The first layer is a stem, which is consisted of a 3D convolutional layer and a PReLU, and the others are 3D residual blocks. The spectral channel is formulated as
\begin{equation}F_{M}^{0}=PReLU(w_{M}^{0}*{{X}_{M}}+b_{M}^{0})\end{equation}
\begin{equation}F_{M,CLSTM}^{l}=R_{3D}^{l}(F_{M}^{l-1})\end{equation}
\begin{equation}F_{M}^{l}=F_{M,CLSTM}^{l}+F_{CLSTM,M}^{l}\end{equation}
\begin{equation}F_{M}^{L}=R_{3D}^{L}(F_{3D}^{L-1})\end{equation}
where $R_{3D}^{l}$ represents the $l$-th 3D residual block, $F_{3D}^{l}$ denotes the output spectral feature at level $l$. The 3D residual block process 3D information using the following formula:
\begin{equation}F_{M}^{l}(x)=x+PReLU(w_{M}^{l,1}*PReLU(w_{M}^{l,0}*x+b_{M}^{l,0})+b_{M}^{l,1})\end{equation}
where $PReLU$ is an activation function.

\subsection{S$^2$-CLSTM module}
Once the spatial and spectral representations $F_{P,CLSTM}^{l}$ and $F_{M,CLSTM}^{l}$ are obtained,  the S$^2$-CLSTM module is utilized to fully integrate these features. The S$^2$-CLSTM module includes two fusion ways: the intra-level fusion and inter-level fusion. The former is carried out via the bi-directional lateral connections and the later via the cell state in the S$^2$-CLSTM.

\begin{figure}[t]
	\setlength{\abovecaptionskip}{-0.5 cm}
	\begin{center}
		\includegraphics[width=0.8\linewidth]{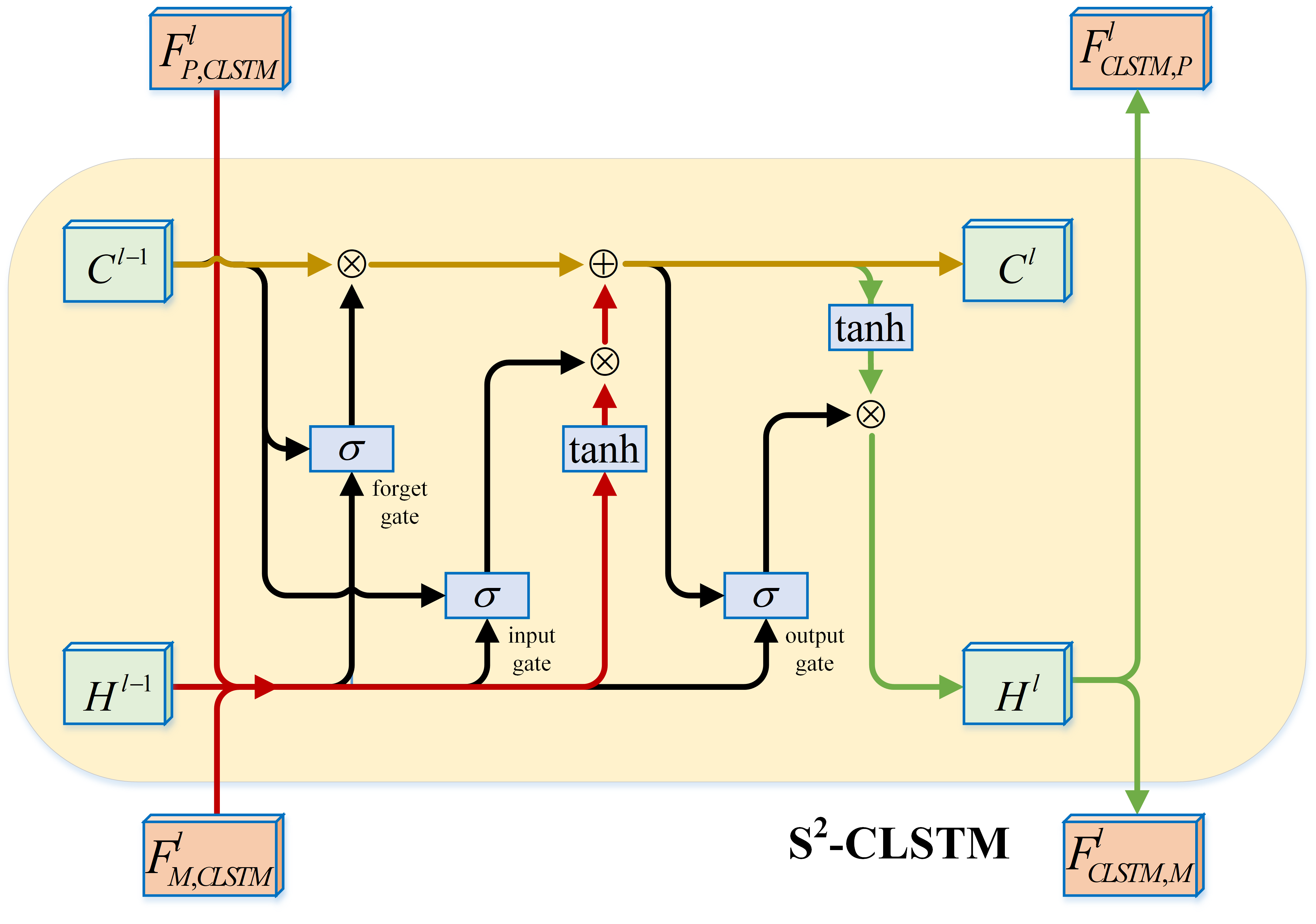}
	\end{center}
	\caption{The arichitecture of S$^2$-CLSTM fusion module. The element-wise multiplication is denoted as $\otimes$. $\oplus$represents the element-wise summation operation. $\sigma$ is the sigmoid activation function, and $\tanh$ indicates the tanh function.}
	\label{fig:S2-CLSTM}
\end{figure}

The architecture of S$^2$-CLSTM is shown in Fig. \ref{fig:S2-CLSTM}. The S$^2$-CLSTM has two inputs ($F_{P,CLSTM}^{l}$ and $F_{M,CLSTM}^{l}$), two outputs ($F_{CLSTM,P}^{l}$ and $F_{CLSTM,M}^{l}$), and three gates (forget gate, input gate, and output gate). $F_{P,CLSTM}^{l}$ and $F_{M,CLSTM}^{l}$  represent the spatial and spectral features of dual channels at the level $l$, where $l$ ranges in $[1,L]$. ${{H}^{l}}$ is the hidden state of S$^2$-CLSTM at the level $l$. ${{C}^{l}}$ indicates the cell state at the level $l$. The hidden state and the cell state are initialized as zero. Each of these gates can be thought of as a "standard" neuron, and they are connected by black lines in Fig. \ref{fig:S2-CLSTM}.

The other three main components of the S$^2$-CLSTM are input flow, output flow, and the cell state. As lateral connections, the input flow (red lines) and output flow (green lines) bi-directionally connect spatial and spectral channels. The cell state (the dark yellow line) integrates features from different levels. Forget gate, input gate, and output gate are used for feature selection for lateral connections and the cell state.

\subsubsection{Intra-level fusion via lateral connections}

In the input flow, the spatial and spectral information from two channels is merged by element-wise addition. The fused spatial-spectral information then passes through the $tanh$ activation function. Finally, the spatial-spectral feature will be selected by the input gate. This process can be formulated as:
\begin{equation}{{i}^{l}}=\sigma({{W}_{P,i}}{{*}^{T}}F_{P,CLSTM}^{l}+{{W}_{M,i}}*F_{M,CLSTM}^{l}+{{W}_{h,i}}*{{H}^{l-1}}+{{W}_{c,i}}*{{C}^{l-1}}+{{b}_{i}})\end{equation}
\begin{equation}F_{CLSTM,i}^{l}={{i}^{l}}\circ\tanh({{W}_{P,c}}*F_{P,CLSTM}^{l}+{{W}_{M,c}}*F_{M,CLSTM}^{l}+{{W}_{h,c}}*{{H}^{l-1}}+{{b}_{c}})\end{equation}
where ${{i}^{l}}$ denotes the activations of input gates at the level $l$, $*$ and ${ * ^T}$ represent the convolution and deconvolution operation, respectively, $W$ and $b$ are learnable parameters of convolutional layers. It is worth noting that the $*$ in ${{W}_{c,i}}*{{C}^{l-1}}$ is a grouped convolution, where the number of groups is the same as the channel dimension. One of the advantages of group convolution is that it can release the restrict of the input image size produced by the Hadamard product in the original CLSTM. Since the size of the input images can be any, the block effect in the pan-sharpened image is eliminated.

The output flow is the information flow from the S$^2$-CLSTM to spatial and spectral channels, which is represented by the green lines in Fig. \ref{fig:S2-CLSTM}. The cell state ${{C}^{l}}$ memories the low-level features. The S$^2$-CLSTM automatically extracts hidden state by output gate ${{o}^{l}}$. Since the spectral channel and S$^2$-CLSTM operate on 3D data, the output feature $F_{CLSTM,P}^{l}$ for the spatial channel needs to be transformed to 2D. The output features can be obtained with the following functions:
\begin{equation}{{o}^{l}}=\sigma({{W}_{P,o}}{{*}^{T}}F_{P,CLSTM}^{l}+{{W}_{M,o}}*F_{M,CLSTM}^{l}+{{W}_{h,o}}*{{H}^{l}}+{{W}_{c,o}}*{{C}^{l}}+{{b}_{o}})
\end{equation}
\begin{equation}{{H}^{l}}={{o}^{l}}\circ \tanh ({{C}^{l}})\end{equation}
\begin{equation}F_{CLSTM,P}^{l}(i,j)=[{{H}^{l}}(i,j,1),\ldots ,{{H}^{l}}(i,j,B)]\end{equation}
\begin{equation}F_{CLSTM,M}^{l}={{H}^{l}}\end{equation}
where $F_{CLSTM,M}^{l}$ is the output feature for the spectral channel, $(i,j)$ indicates the spatial location of the pixel in feature maps, and$[\cdot ]$is a concatenation operation.

\subsubsection{Inter-level fusion via the cell state}

As the fused spatial-spectral features of previous levels are memorized in the cell state ${{C}^{l-1}}$, ${{C}^{l-1}}$ acts as the bridge to connect the current level with previous levels, which can effectivley capture the underlying correslations among mult-level representations and facilitate inter-level fusion in different levels. Dark yellow lines show the information flow of inter-level fusion in Fig. \ref{fig:S2-CLSTM}. The following equations are presented for this procedure:
\begin{equation}{{f}^{l}}=\sigma({{W}_{P,f}}{{*}^{T}}F_{P}^{l}+{{W}_{M,f}}*F_{M}^{l}+{{W}_{h,f}}*{{H}^{l-1}}+{{W}_{c,f}}*{{C}^{l-1}}+{{b}_{f}})\end{equation}
\begin{equation}{{C}^{l}}={{f}^{l}}\circ {{C}^{l-1}}+F_{CLSTM,i}^{l}\end{equation}
where the little circle $\circ$ represents an element-wise multiplication, $W$ and $b$ are parameters of convolutional layers.

\subsection{Reconstruction module}

\begin{figure}[t]
	\setlength{\abovecaptionskip}{-0.5 cm}
	\begin{center}
		\includegraphics[width=0.8\linewidth]{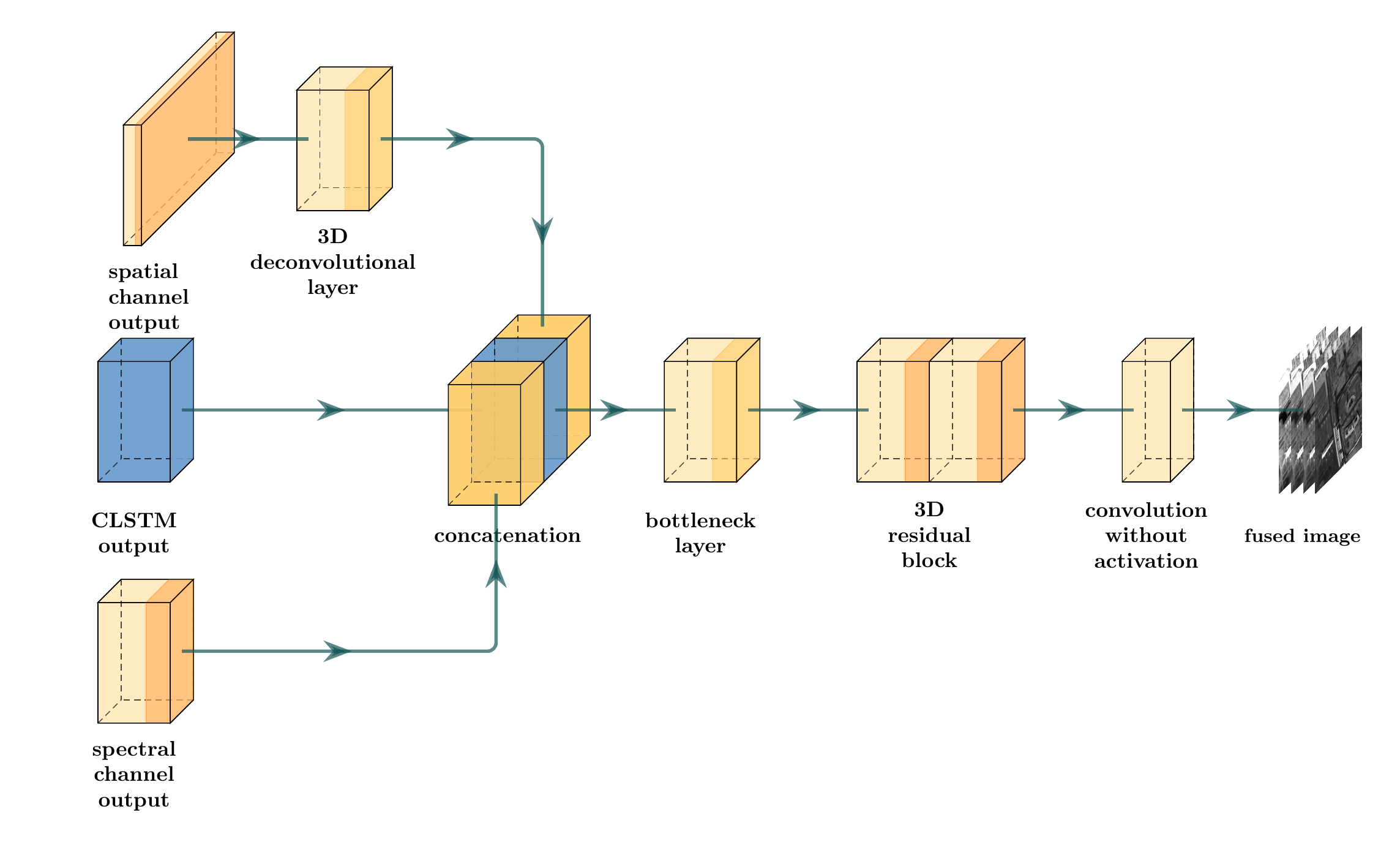}
	\end{center}
	\caption{The architecture of the reconstruction module}
	\label{fig:reconstruction}
\end{figure}

As the last part of the proposed DCNet, the reconstruction module will recover the desired HR-MS image from the output features $F_P^L$,$F_M^L$, and $H^L$ of the spatial channel, the spectral channel, and S$^2$-CLSTM, respectively. Fig. \ref{fig:reconstruction} illustrates the reconstruction module, which can be divided into four components: a 3D de-convolutional layer, a bottleneck layer, a 3D residual block, and a 3D convolutional layer without activation. First, the feature from the spatial channel is projected into $R^{\beta C\times B\times H\times W}$ by a de-convolutional layer. Next, we concatenate it with $F_M^L$ and $H^L$. Then, the bottleneck layer is added to weight the three 3D features by $\beta C$ filters of size $1\times1\times1$. After that, the output of this layer is fed into a 3D residual block $R_{3D}$. Finally, $B$ filters of size $3\times3\times3$ in the convolutional layer will recover the ideal HR-MS image.

\subsection{Objective function}

In the training phase, given the DCNet $\Phi (\cdot ,\theta )$ parameterized by $\theta $, the objective is to find $\theta$. The training set ${{X}_{train}}$ has $N$ pairs of ${{x}_{P}}$ (PAN image), ${{x}_{M}}$(MS image), and ${{y}_{M}}$ (ground truth). Accordingly, the object function can be formulated as:
\begin{equation}
\theta=\underset{\theta}{\arg \min } J\left(X_ {train}, \theta\right) \triangleq \underset{\theta}{\arg \min } \frac{1}{N} \sum_{y_{M, n} \in X_{tain}} L\left(\hat{y}_{M, n}, y_{M, n}\right)
\end{equation}
\begin{equation}
L\left(\hat{y}_{M}, y_{M}\right)=L\left(\Phi\left(x_{P}, x_{M}, \theta\right), y_{M}\right)=\left\|\Phi\left(x_{P}, x_{M}, \theta\right)-y_{M}\right\|_{1}+\lambda\|\theta\|_{2}^{2}
\end{equation}
where $L(\cdot)$ is a loss function. The first part of the loss function is a l1-loss function, which is efficient and edge-sensitive. To prevent overfitting, we add the l2 penalty $\left\| \theta  \right\|_{2}^{2}$ as a regularization item in the loss function. The $\lambda$ is a balancing parameter that balances the importance of the l1-loss and the regularization term. Upon convergence, the parameter $\theta$ is frozen and can be used for tests on both the lower scale and the original scale data.

\section{Experiments}
\label{Experiments}

\subsection{Data sets}

In this section, three datasets are constructed to compare the performance of DCNet with state-of-art networks. The original date is acquired by three satellites: IKONOS, GaoFen-2, and WorldView-2. Each satellite carries a PAN sensor and an MS sensor. The details of these datasets are demonstrated below.

The first dataset was acquired by the IKONOS satellite over the mountainous area in the west of Sichuan Province, China, in 2008. The spectral resolutions of PAN, blue, green, red, and near-infrared bands are $450-900$ nm, $~450-530$ nm, $520-610$ nm, $640-720$ nm, and $76-60$ nm, respectively. We got two pairs of PAN and MS images. The PAN images consist of ${12096}\times{11408}$ and ${11552}\times {18112}$ pixels, respectively, while the spatial resolution of MS images is ${3024}\times {2852}$ and ${2888}\times {4528}$. The spatial resolutions of these PAN and MS images are 4 and 1 m.

The second dataset was taken over the Guangzhou, China mall by GaoFen-2 in 2016 with PAN and MS dimensions are ${23552}\times {20080}$ and ${5888}\times {5020}$ respectively. The spectral resolutions of PAN, blue, green, red, and near-infrared bands are $450-900$, $~450-520$, $520-590$, $630-690$, and $770-890$ nm, respectively. The spatial resolutions are the same as IKONOS.

The third dataset was acquired by the WorldView-2 satellite over Washington, DC, USA, in 2016. The spectral resolutions of Coastal, Blue, Green, yellow, red, Red edge, Near-IR1, Near-IR2, and bands are $400-450$, $450-510$, $510-580$, $585-625$, $630-690$, $705-745$, $770-895$, $860-1040$, and $450-800$ nm, respectively. We got one pair of PAN and MS images. The PAN images consist of ${24,352}\times {24,922}$ pixels, respectively, while the spatial resolution of MS images is ${6088}\times {6248}$. Different from IKONOS and GaoFen-2, the WorldView-2 has a higher spatial resolution 0.5 and 2 m for PAN and MS images. 

\begin{table}[!htbp]
	\caption{The distribution of images for training, validation, and testing.}
	\centering
	\begin{tabular}{c c c c}
		\hline
		Dataset & Train set & Validation set & Test set \\
		\hline
		IKONOS & 400 & 100 & 50 \\
		GaoFen-2 & 400 & 100 & 50 \\
		WorldView-2 & 400 & 100 & 50\\
		\hline
	\end{tabular}
	\label{tab:dataset_distribution}
\end{table}

From the above, we can find that the spectral resolution of IKONOS, GaoFen-2, and WorldView-2 differs from each other. Thus, the datasets cannot be fused for training and testing. The remote sensing images mentioned above are cut into patches. Because of the volume of our network, the height and width of the train set and validation set are 128. But the test set block has a larger height and width, which is 256. For a better test of the generalization, the resulting datasets are covering different types of areas, e.g., urban, mountain, lake, and so on. 1,650 images were collected in total, 550 images for each dataset. 

Following \cite{huang2015new}, original images with 4 or 8 bands from IKONOS, GaoFen-2, and WorldView-2 were used as the ground truth , and it was down-sampled using bicubic interpolation algorithm to obtain the simulated MS images   with low spatial resolution according to Wald’s protocol \cite{Wald1997Fusion}. Meanwhile, we have down-sampled PAN images   using the same process. The distribution of images of the resulting datasets is listed in Table \ref{tab:dataset_distribution}.

\subsection{Experiment setting}

\begin{table}
	\caption{Hyper-parameters of the DCNet for the IKONOS and GaoFen-2 datasets.}
	\centering
	\begin{tabular}{ c c c c c}
		\hline
		Channel &  & Stem &Level 1, 2, and 3 & Level 4\\
		\hline
		\multirow{4}{*}{\makecell[c]{Spatial\\channel}} & Kernel/Stride & \makecell[c]{${{3}^{2}}\times 128/1$}& \makecell[c]{${{3}^{2}}\times 128/1$}& \makecell[c]{${{3}^{2}}\times 128/1$}\\
		& Input & \makecell[c]{${{128}^{2}}$}& \makecell[c]{${{128}^{2}}\times 128$}& \makecell[c]{${{128}^{2}}\times 128$} \\
		& $F_P$ & \makecell[c]{${{128}^{2}}\times 128$}& \makecell[c]{${{128}^{2}}\times 128$}& - \\
		& $F_{P,CLSTM}$ & - & \makecell[c]{${{128}^{2}}\times 4\times 32$}& \makecell[c]{${{128}^{2}}\times 4\times 32$} \\
		\cline{2-5}
		\multirow{4}{*}{\makecell[c]{Spectral\\channel}} & Kernel/Stride & \makecell[c]{${{3}^{2}}\times 4\times 32/1$}& \makecell[c]{${{3}^{2}}\times 4\times 32/1$}& \makecell[c]{${{3}^{2}}\times 4\times 32/1$} \\
		& Input & \makecell[c]{${{128}^{2}}\times 4$}& \makecell[c]{${{128}^{2}}\times 4\times 32$}& \makecell[c]{${{128}^{2}}\times 4\times 32$} \\
		& $F_M$ & \makecell[c]{${{128}^{2}}\times 4\times 32$}& \makecell[c]{${{128}^{2}}\times 4\times 32$}& - \\
		& $F_{M,CLSTM}$ & - & \makecell[c]{${{128}^{2}}\times 4\times 32$}& \makecell[c]{${{128}^{2}}\times 4\times 32$} \\
		\hline
	\end{tabular}
	
	\label{tab:hyper-parameters-IKONOS-GaoFen-2}
\end{table}

\begin{table}
	\caption{Hyper-parameters of the DCNet for the WorldView-2 dataset.}
	\centering
	\begin{tabular}{ c c c c c}
		\hline
		Channel &  & Stem &Level 1, 2, and 3 & Level 4\\
		\hline
		\multirow{4}{*}{\makecell[c]{Spatial\\channel}} & Kernel/Stride & \makecell[c]{${{3}^{2}}\times 128/1$}& \makecell[c]{${{3}^{2}}\times 128/1$}& \makecell[c]{${{3}^{2}}\times 128/1$}\\
		& Input & \makecell[c]{${{128}^{2}}$}& \makecell[c]{${{128}^{2}}\times 128$}& \makecell[c]{${{128}^{2}}\times 128$} \\
		& $F_P$ & \makecell[c]{${{128}^{2}}\times 128$}& \makecell[c]{${{128}^{2}}\times 128$}& - \\
		& $F_{P,CLSTM}$ & - & \makecell[c]{${{128}^{2}}\times 4\times 32$}& \makecell[c]{${{128}^{2}}\times 4\times 32$} \\
		\cline{2-5}
		\multirow{4}{*}{\makecell[c]{Spectral\\channel}} & Kernel/Stride & \makecell[c]{${{3}^{2}}\times 8\times 16/1$}& \makecell[c]{${{3}^{2}}\times 8\times 16/1$}& \makecell[c]{${{3}^{2}}\times 8\times 16/1$} \\
		& Input & \makecell[c]{${{128}^{2}}\times 8$}& \makecell[c]{${{128}^{2}}\times 8\times 16$}& \makecell[c]{${{128}^{2}}\times 8\times 16$} \\
		& $F_M$ & \makecell[c]{${{128}^{2}}\times 8\times 16$}& \makecell[c]{${{128}^{2}}\times 8\times 16$}& - \\
		& $F_{M,CLSTM}$ & - & \makecell[c]{${{128}^{2}}\times 8\times 16$}& \makecell[c]{${{128}^{2}}\times 8\times 16$} \\
		\hline
	\end{tabular}
	
	\label{tab:hyper-parameters-WorldView-2}
\end{table}

We implemented the proposed network using the PyTorch framework \cite{ketkar2017introduction}. For each dataset, the proposed model was trained for 1600 epochs over the entire dataset, and we selected Adam \cite{kingma2014adam} to train the proposed network. The experiments were carried out on a GPU server. Two NVIDIA GeForce TITAN Xp GPUs (12GB memory per GPU) are used for training. The batch size was set to 20. The learning rate was initially set to 0.001 and reduced 20\% per 150 epochs. The other hyper-parameters of DCNet are shown below in Tables \ref{tab:hyper-parameters-IKONOS-GaoFen-2} and \ref{tab:hyper-parameters-WorldView-2}. The kernel dimensions of the spatial channel are denoted by ${{(W)}^{2}}\times C/S$ for width, channel, and stride sizes. In the spectral channel, the kernels and strides are represented as ${{W}^{2}}\times B\times C/S$, where $B$ indicates the number of bands. The representation of spatial and spectral features takes the form of ${{W}^{2}}\times C$ and ${{W}^{2}}\times B\times C$, respectively.

\subsection{Evaluation at lower scale}

As mentioned above, training samples come from IKONOS, GaoFen-2, and WorldView-2 satellites. In this section, the proposed DCNet is compared with six state-of-the-art methods including: PNN \cite{masi2016pansharpening}, PNN+ \cite{scarpa2018target}, DRPNN \cite{wei2017boosting}, PanNet \cite{yang2017pannet}, MSDCNN \cite{yuan2018multiscale}, and ResTFNet \cite{liu2020remote}. We assess these methods by visual evaluation and quantitative evaluation by evaluation metrics.

\begin{figure}
	\setlength{\abovecaptionskip}{0.cm}
	\centering
	\subfigure [Ground truth]{
		\includegraphics[width=0.22\linewidth]{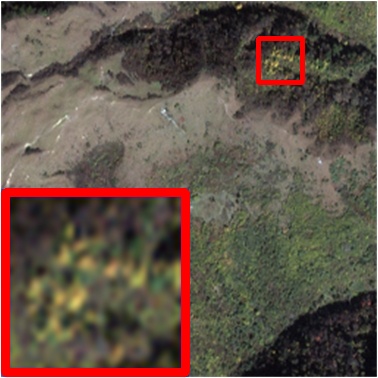}
		\label{fig:visualization_of_IKONOS_GT}
	}
	\subfigure [DCNet(ours)]{
		\includegraphics[width=0.22\linewidth]{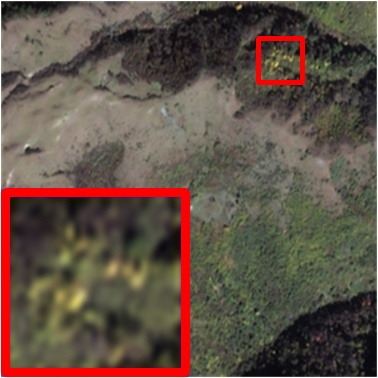}
		\label{fig:visualization_of_IKONOS_DCNet}
	}
	\subfigure [ResTFNet]{
		\includegraphics[width=0.22\linewidth]{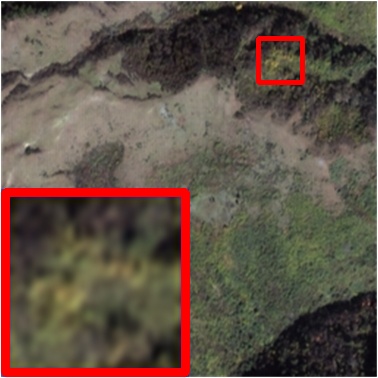}
		\label{fig:visualization_of_IKONOS_ResTFNet}
	}
	\subfigure [MSDCNN]{
		\includegraphics[width=0.22\linewidth]{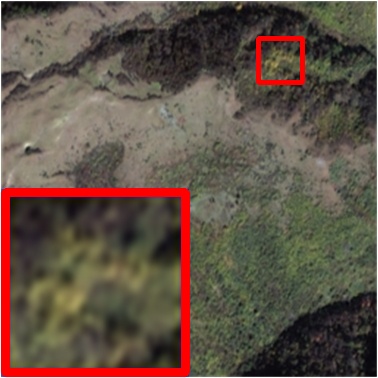}
		\label{fig:visualization_of_IKONOS_MSDCNN}
	}
	
	\subfigure [PNN]{
		\includegraphics[width=0.22\linewidth]{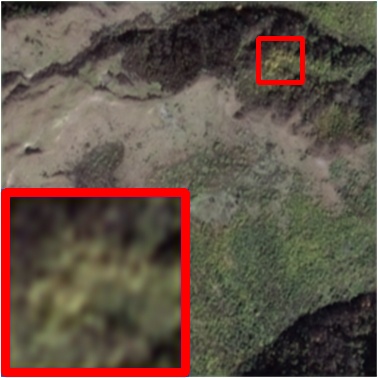}
		\label{fig:visualization_of_IKONOS_PNN}
	}
	\subfigure [PanNet]{
		\includegraphics[width=0.22\linewidth]{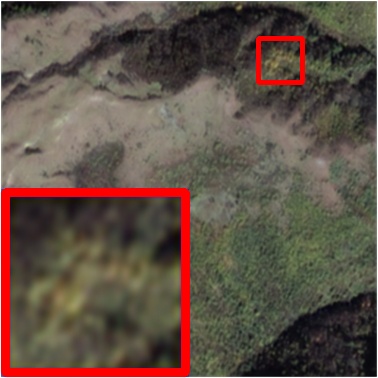}
		\label{fig:visualization_of_IKONOS_PanNet}
	}
	\subfigure [PNN+]{
		\includegraphics[width=0.22\linewidth]{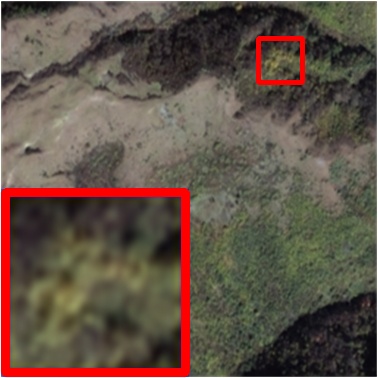}
		\label{fig:visualization_of_IKONOS_PNN+}
	}
	\subfigure [DRPNN]{
		\includegraphics[width=0.22\linewidth]{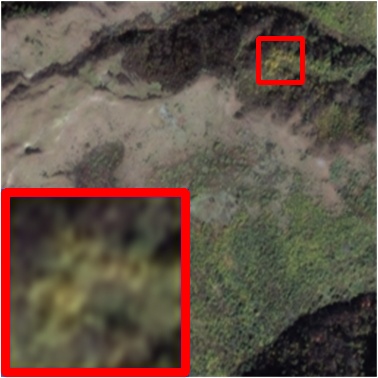}
		\label{fig:visualization_of_IKONOS_DRPNN}
	}
	\caption{Pan-sharpened images by different methods on the IKONOS dataset.}
	\label{fig:visualization_of_IKONOS}
\end{figure}

For \textbf{visual evaluation}, the fused images are visualized to check spatial and spectral distortions. We begin with the IKONOS dataset. Fig. \ref{fig:visualization_of_IKONOS} shows an example of the experiment performs on an IKONOS image. Since MS images have more than three bands, only red, green, and blue bands are extracted to synthesize the TrueColor images. The ground truth is shown in Fig. \ref{fig:visualization_of_IKONOS_GT}. Fig. \ref{fig:visualization_of_IKONOS_DCNet}-(h) display the pan-sharpened images by different methods. The proposed DCNet produces the pan-sharpened image with the best visual quality in terms of spectral preservation, e.g., the yellow part reconstructed by the proposed network is most close to the ground truth. The proposed DCNet does better in spectral preservation and provides images with more precious spatial details.

\begin{figure}
	\setlength{\abovecaptionskip}{0.cm}
	\centering
	\subfigure [Ground truth]{
		\includegraphics[width=0.22\linewidth]{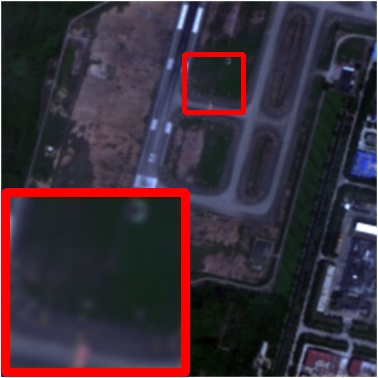}
		\label{fig:visualization_of_GaoFen-2_GT}
	}
	\subfigure [DCNet(ours)]{
		\includegraphics[width=0.22\linewidth]{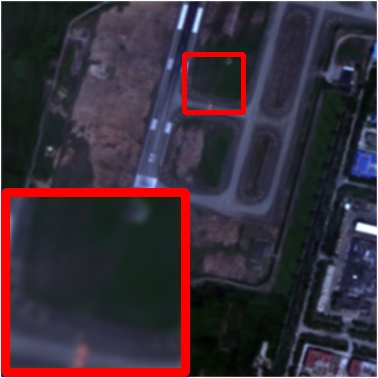}
		\label{fig:visualization_of_GaoFen-2_DCNet}
	}
	\subfigure [ResTFNet]{
		\includegraphics[width=0.22\linewidth]{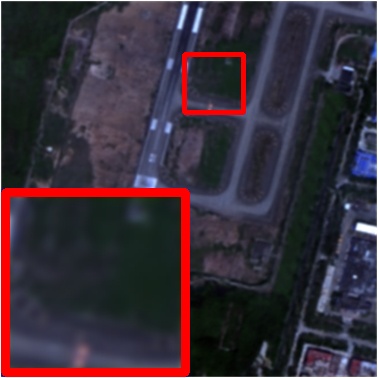}
		\label{fig:visualization_of_GaoFen-2_ResTFNet}
	}
	\subfigure [MSDCNN]{
		\includegraphics[width=0.22\linewidth]{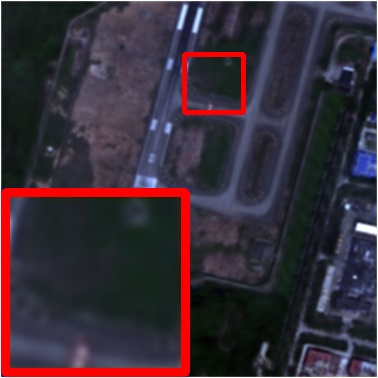}
		\label{fig:visualization_of_GaoFen-2_MSDCNN}
	}
	
	\subfigure [PNN]{
		\includegraphics[width=0.22\linewidth]{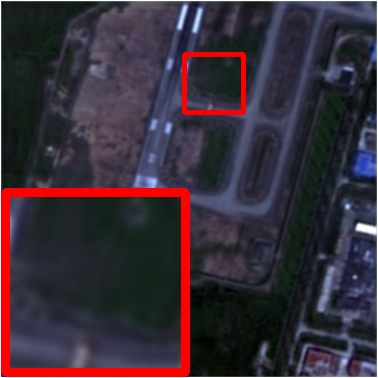}
		\label{fig:visualization_of_GaoFen-2_PNN}
	}
	\subfigure [PanNet]{
		\includegraphics[width=0.22\linewidth]{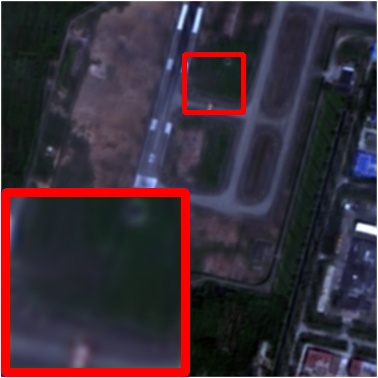}
		\label{fig:visualization_of_GaoFen-2_PanNet}
	}
	\subfigure [PNN+]{
		\includegraphics[width=0.22\linewidth]{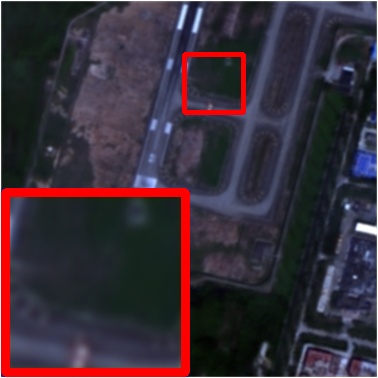}
		\label{fig:visualization_of_GaoFen-2_PNN+}
	}
	\subfigure [DRPNN]{
		\includegraphics[width=0.22\linewidth]{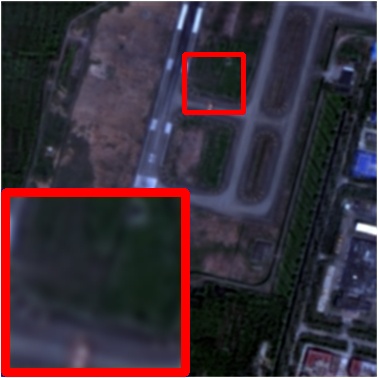}
		\label{fig:visualization_of_GaoFen-2_DRPNN}
	}
	\caption{Pan-sharpened images by different methods on the GaoFen-2 dataset.}
	\label{fig:visualization_of_GaoFen-2}
\end{figure}

\begin{figure}
	\setlength{\abovecaptionskip}{0.cm}
	\centering
	\subfigure [Ground truth]{
		\includegraphics[width=0.22\linewidth]{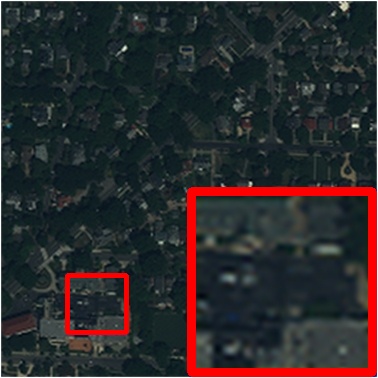}
		\label{fig:visualization_of_WorldView-2_GT}
	}
	\subfigure [DCNet(ours)]{
		\includegraphics[width=0.22\linewidth]{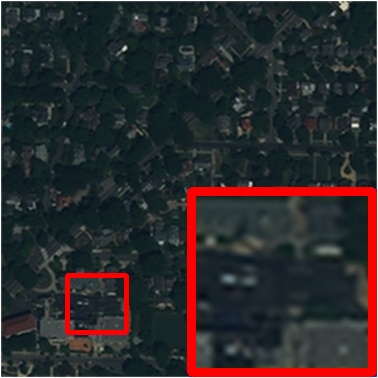}
		\label{fig:visualization_of_WorldView-2_DCNet}
	}
	\subfigure [ResTFNet]{
		\includegraphics[width=0.22\linewidth]{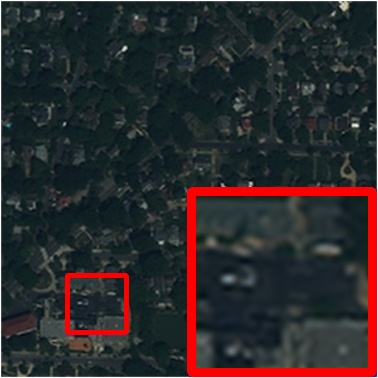}
		\label{fig:visualization_of_WorldView-2_ResTFNet}
	}
	\subfigure [MSDCNN]{
		\includegraphics[width=0.22\linewidth]{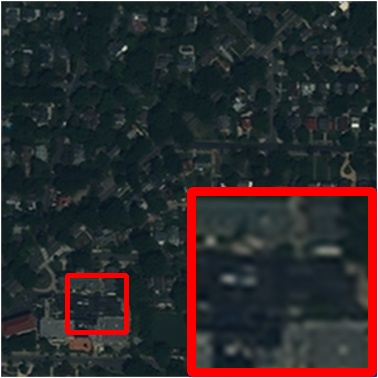}
		\label{fig:visualization_of_WorldView-2_MSDCNN}
	}
	
	\subfigure [PNN]{
		\includegraphics[width=0.22\linewidth]{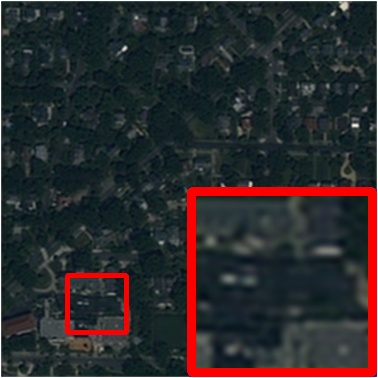}
		\label{fig:visualization_of_WorldView-2_PNN}
	}
	\subfigure [PanNet]{
		\includegraphics[width=0.22\linewidth]{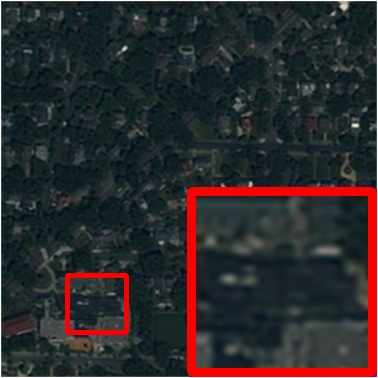}
		\label{fig:visualization_of_WorldView-2_PanNet}
	}
	\subfigure [PNN+]{
		\includegraphics[width=0.22\linewidth]{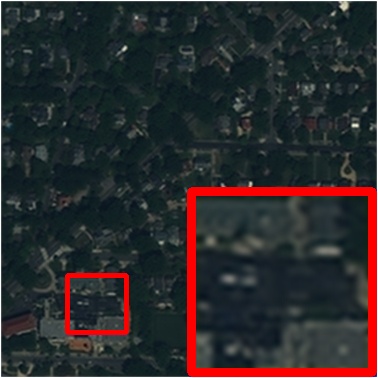}
		\label{fig:visualization_of_WorldView-2_PNN+}
	}
	\subfigure [DRPNN]{
		\includegraphics[width=0.22\linewidth]{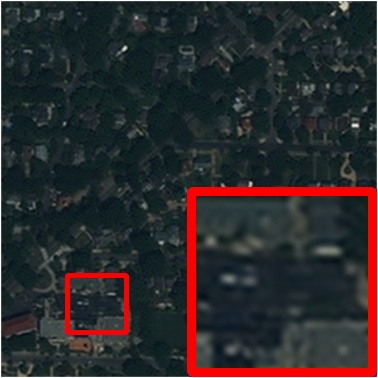}
		\label{fig:visualization_of_WorldView-2_DRPNN}
	}
	\caption{Pan-sharpened images by different methods on the WorldView-2 dataset.}
	\label{fig:visualization_of_WorldView-2}
\end{figure}

Fig. \ref{fig:visualization_of_GaoFen-2} illustrates an experiment performed on a GaoFen-2 image. We have similar observations in Fig. \ref{fig:visualization_of_IKONOS}. PNN, PNN+, DRPNN, PanNet, and MSDCNN produce spectral distortion, which is indicated by the white rectangle object that is pink in the reference MS image. TFNet can retrain most spectral information of the pink rectangle object, but the lunar shape object changes to an L shape. The fused image produced by LGC is blurred. The proposed network can obtain the closest pan-sharpened image to the ground truth no matter at the white lunar shape object nor the pink rectangle. The DCNet produces lest distortion in both spatial and spectral domains. Although it cannot identify the apparent difference between the pan-sharpened images in Fig. \ref{fig:visualization_of_WorldView-2}, the quantitative evaluation in Table 6 demonstrates the excellent performance of the proposed DCNet.

To \textbf{quantitatively evaluate} the performance of the DCNet and state-of-art methods, five popular indices have been employed. They are Q4 (for 4-band) or Q8 (for 8-band) \cite{zeng2010fusion}, universal image quality index (UIQI) \cite{wang2002universal}, spectral angle mapper (SAM) \cite{dennison2004comparison}, relative dimensionless global error in synthesis (ERGAS) \cite{ayhan2012spectral} and spatial correlation coefficient (SCC) \cite{zhou1998wavelet}. Q4/Q8, UIQI, and ERGAS can comprehensively evaluate the spectral and spatial quality of the fused image. SCC is a widely used index to measure the spatial quality of the fused image. In contrast, SAM can effectively measure spectral distortion in the fused image compared with the reference image.

\begin{table}
	\caption{Quantitative evaluation results of different methods on the IKONOS dataset. The optimal and the sub-optimal results are in \textcolor{red}{red} and \textcolor{blue}{blue}, respectively.}
	\centering
	\begin{tabular}{ c c c c c c}
		\hline
		Method & Q4 & UIQI & SAM & ERGAS & SCC\\
		\hline
		PNN \cite{masi2016pansharpening} & 0.5474 &	0.8458 & 4.9105 & 4.2016 & 0.8010\\
		PNN+ \cite{scarpa2018target}&0.5518&0.8599&4.2989&3.9579&0.8173\\
		DRPNN \cite{wei2017boosting}&0.5320&0.8465&4.5075&4.1639&0.8055\\
		PanNet \cite{yang2017pannet}&0.5293&0.8393&4.7777&4.2836&0.8014\\
		MSDCNN \cite{yuan2018multiscale}&0.5683&0.8686&4.1819&3.7883&0.8328\\
		ResTFNet \cite{liu2020remote}&\textcolor{blue}{0.6592}&\textcolor{blue}{0.8964}&\textcolor{blue}{4.1303}&\textcolor{blue}{3.3761}&\textcolor{blue}{0.9186}\\
		DCNet(ours)&\textcolor{red}{0.7236}&\textcolor{red}{0.9242}&\textcolor{red}{3.6596}&\textcolor{red}{2.7353}&\textcolor{red}{0.9502}\\
		\hline
		Ideal value&1&1&0&0&1\\
		\hline
	\end{tabular}
	
	\label{tab:results-low-resolution-IKONOS}
\end{table}

\begin{table}
	\caption{Quantitative evaluation results of different methods on the GaoFen-2 dataset. The optimal and the sub-optimal results are in \textcolor{red}{red} and \textcolor{blue}{blue}, respectively.}
	\centering
	\begin{tabular}{ c c c c c c}
		\hline
		Method & Q4 & UIQI & SAM & ERGAS & SCC\\
		\hline
		PNN \cite{masi2016pansharpening} &0.7390&0.9094&4.3010&4.3037&	0.8776\\
		PNN+ \cite{scarpa2018target}&0.7843&0.9488&3.0594&3.5287&	0.9092\\
		DRPNN \cite{wei2017boosting}&0.7275&0.8719&4.6670&4.8960&	0.8515\\
		PanNet \cite{yang2017pannet}&0.7504&0.9008&4.1774&4.4150&	0.8752\\
		MSDCNN \cite{yuan2018multiscale}&\textcolor{blue}{0.8031}&\textcolor{blue}{0.9507}&\textcolor{blue}{3.0032}&\textcolor{blue}{3.2756}&\textcolor{blue}{0.9213}\\
		ResTFNet \cite{liu2020remote}&0.6907&0.9164&3.7689&3.9698&	0.9139\\
		DCNet(ours)&\textcolor{red}{0.8707}&\textcolor{red}{0.9741}&\textcolor{red}{2.1171}&\textcolor{red}{2.3805}&\textcolor{red}{0.9655}\\
		\hline
		Ideal value&1&1&0&0&1\\
		\hline
	\end{tabular}
	
	\label{tab:results-low-resolution-GaoFen-2}
\end{table}

\begin{table}
	\caption{Quantitative evaluation results of different methods on the WorldView-2 dataset. The optimal and the sub-optimal results are in \textcolor{red}{red} and \textcolor{blue}{blue}, respectively.}
	\centering
	\begin{tabular}{ c c c c c c}
		\hline
		Method & Q4 & UIQI & SAM & ERGAS & SCC\\
		\hline
		PNN \cite{masi2016pansharpening} & 0.5520&0.8717&7.0039&4.1212& 0.8709\\
		PNN+ \cite{scarpa2018target}&0.6209&0.9027&5.6871&3.4700&0.9116\\
		DRPNN \cite{wei2017boosting}&0.6399&0.9025&5.6503&3.3753&0.9226\\
		PanNet \cite{yang2017pannet}&0.5604&0.8740&6.9058&4.1172&0.8742\\
		MSDCNN \cite{yuan2018multiscale}&\textcolor{blue}{0.6628}&\textcolor{blue}{0.9124}&\textcolor{blue}{5.3072}&\textcolor{blue}{3.1768}&\textcolor{blue}{0.9303}\\
		ResTFNet \cite{liu2020remote}&0.6587&0.9107&5.3367&3.3563&	0.9180\\
		DCNet(ours)&\textcolor{red}{0.6982}&\textcolor{red}{0.9249}&\textcolor{red}{4.6668}&\textcolor{red}{2.8387}&\textcolor{red}{0.9476}\\
		\hline
		Ideal value&1&1&0&0&1\\
		\hline
	\end{tabular}
	
	\label{tab:results-low-resolution-WorldView-2}
\end{table}

The quantitative evaluation results are shown in Tables \ref{tab:results-low-resolution-IKONOS}-\ref{tab:results-low-resolution-WorldView-2}. The optimal and sub-optimal results are shown in \textcolor{red}{red} and \textcolor{blue}{blue}, respectively. For the spectral metric SAM, the spatial metric SCC, or other global metrics, DCNet substantially outperforms other methods. Specifically, for the SAM index on the GaoFen-2 dataset, DCNet achieves 2.1171 even though the sub-optimal result is 3.0032. It can demonstrate that the proposed DCNet significantly exceeds all other state-of-the-art methods, and the pan-sharpened images by DCNet have lest spatial and spectral distortion.

\subsection{Evaluation at original scale}

In this section, we will compare the proposed DCNet with state-of-the-art methods at the original scale of PAN and MS images. As there are no ground truth images, we use the model trained on lower scale images, and the original PAN and MS images are adopted as the spatial and spectral references, respectively. PNN PNN+, DRPNN, PanNet, MSDCNN, ResTFNet, and the proposed network are evaluated and compared by both visual evaluation and quantitative evaluation.

\begin{figure}
	\setlength{\abovecaptionskip}{0 cm}
	\centering
	\subfigure [PAN]{
		\includegraphics[width=0.3\linewidth]{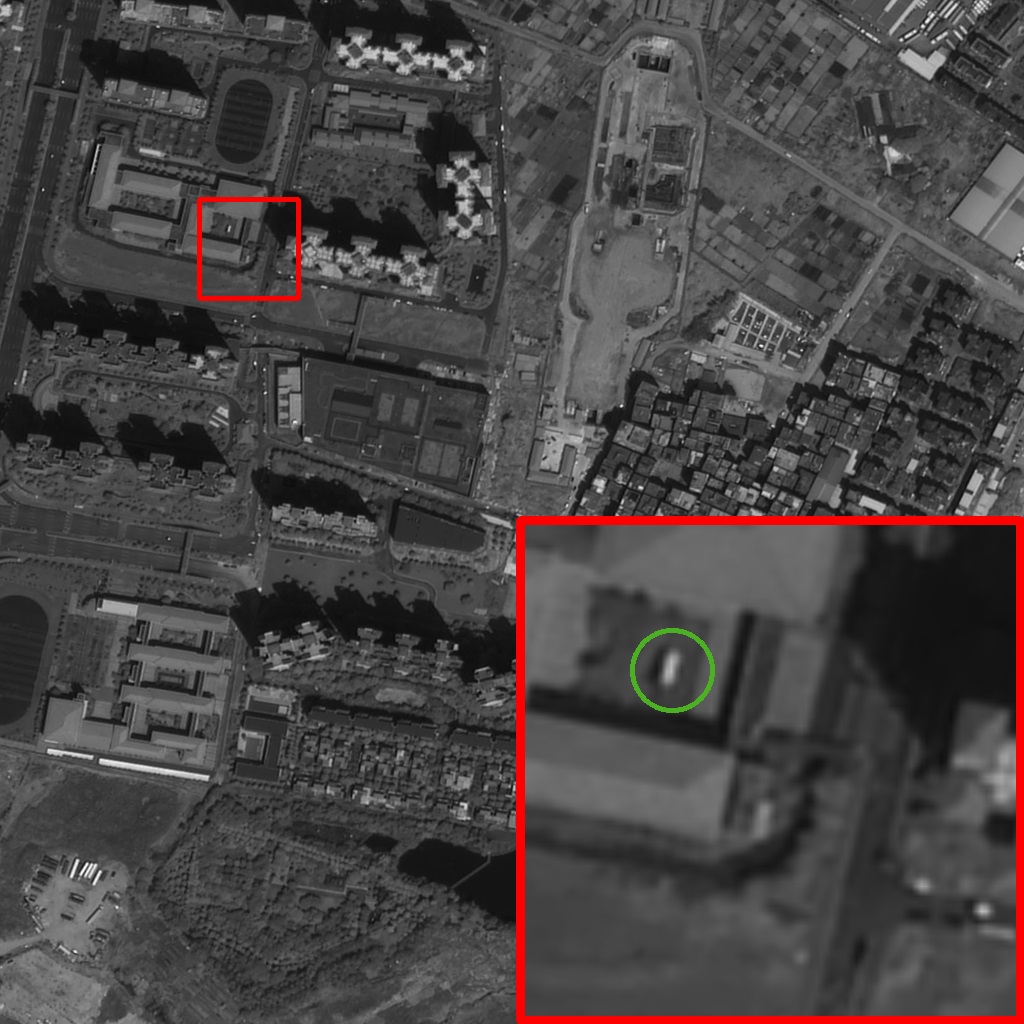}
		\label{fig:visualization_of_original_scale_PAN}
	}
	\subfigure [MS]{
		\includegraphics[width=0.3\linewidth]{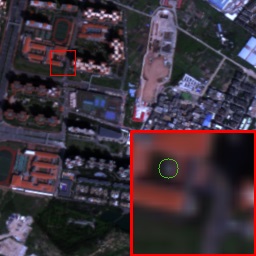}
		\label{fig:visualization_of_original_scale_MS}
	}
	\subfigure [DCNet(ours)]{
		\includegraphics[width=0.3\linewidth]{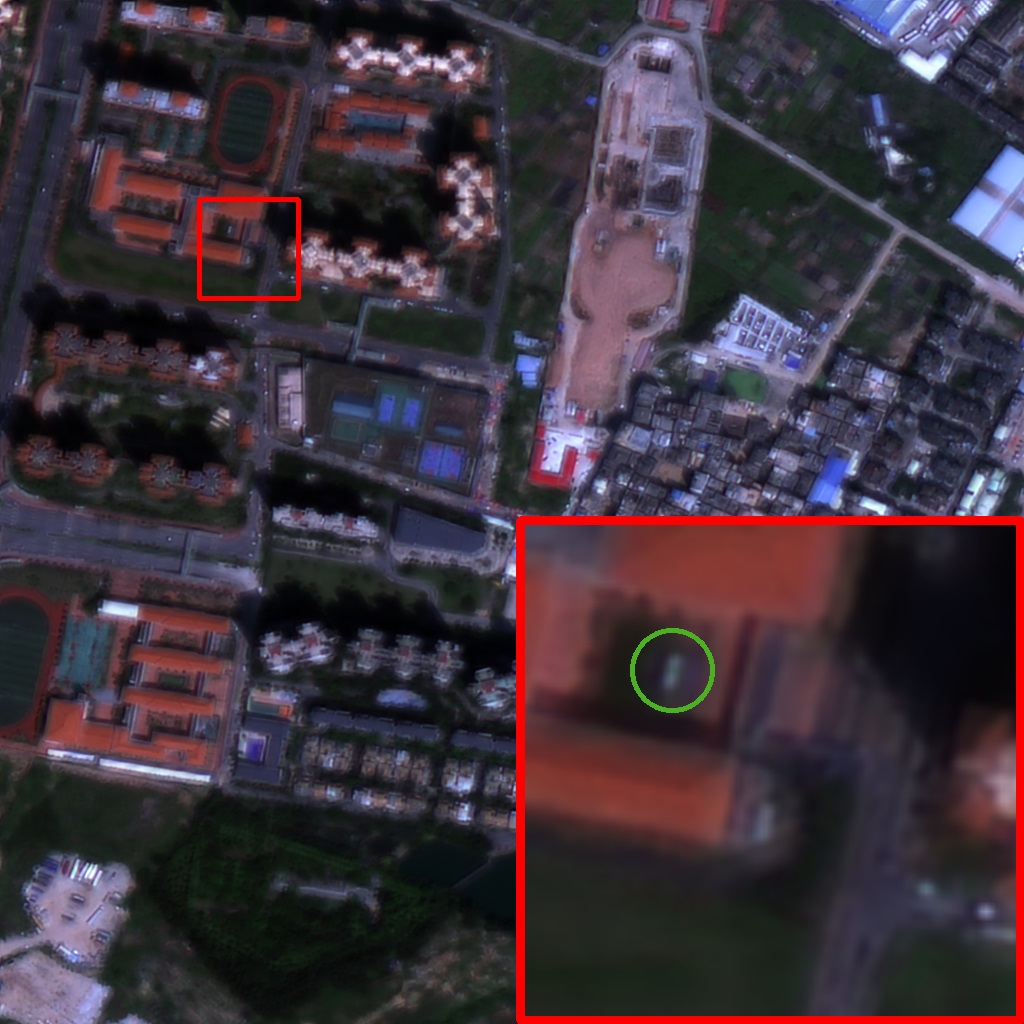}
		\label{fig:visualization_of_original_scale_DCNet}
	}
	\subfigure [ResTFNet]{
		\includegraphics[width=0.3\linewidth]{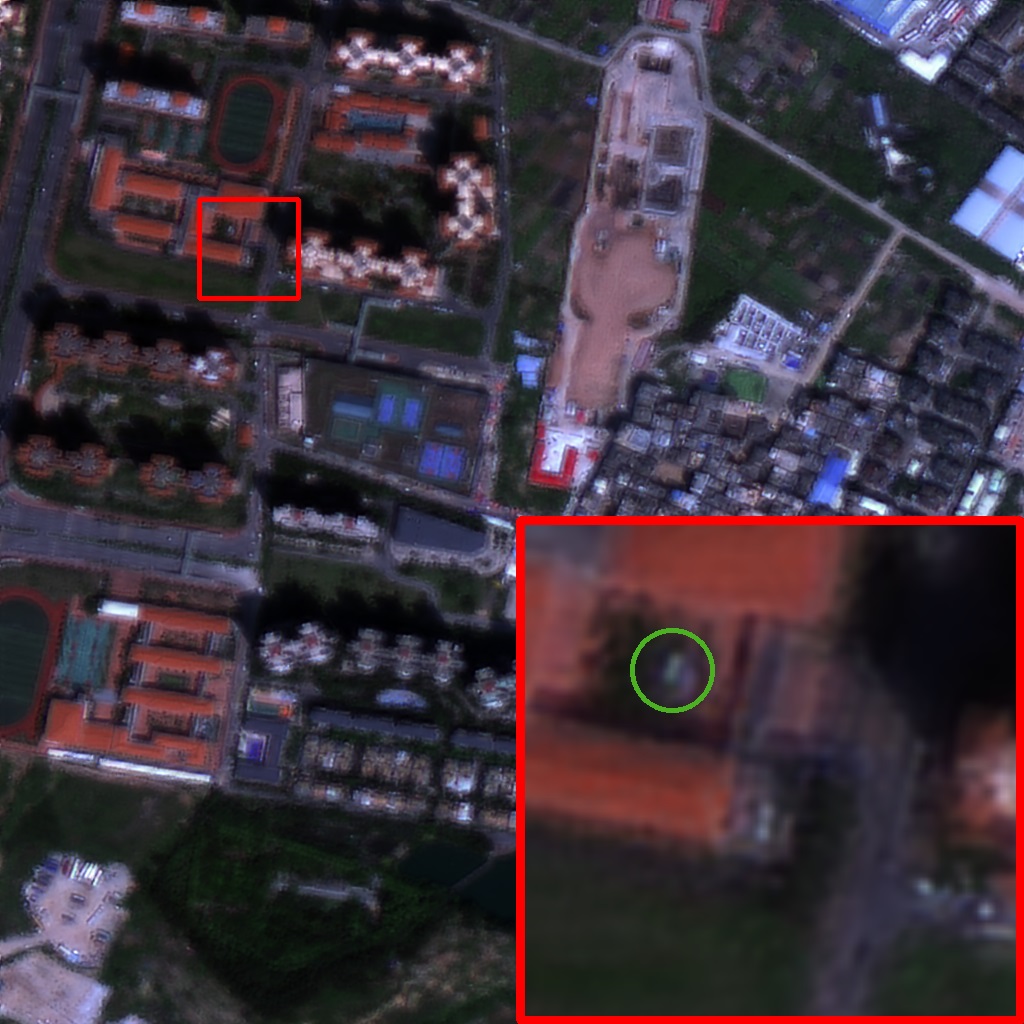}
		\label{fig:visualization_of_original_scale_ResTFNet}
	}
	\subfigure [MSDCNN]{
		\includegraphics[width=0.3\linewidth]{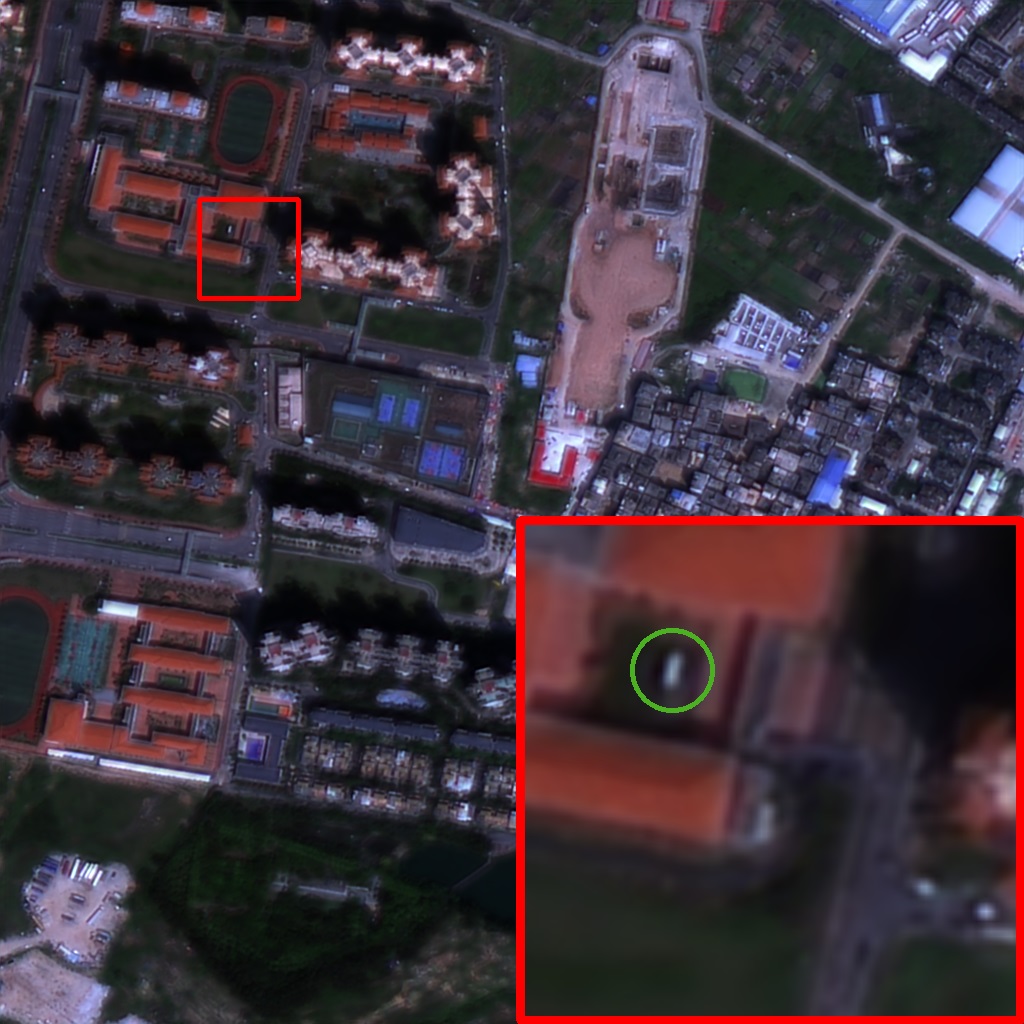}
		\label{fig:visualization_of_original_scale_MSDCNN}
	}
	\subfigure [PNN]{
		\includegraphics[width=0.3\linewidth]{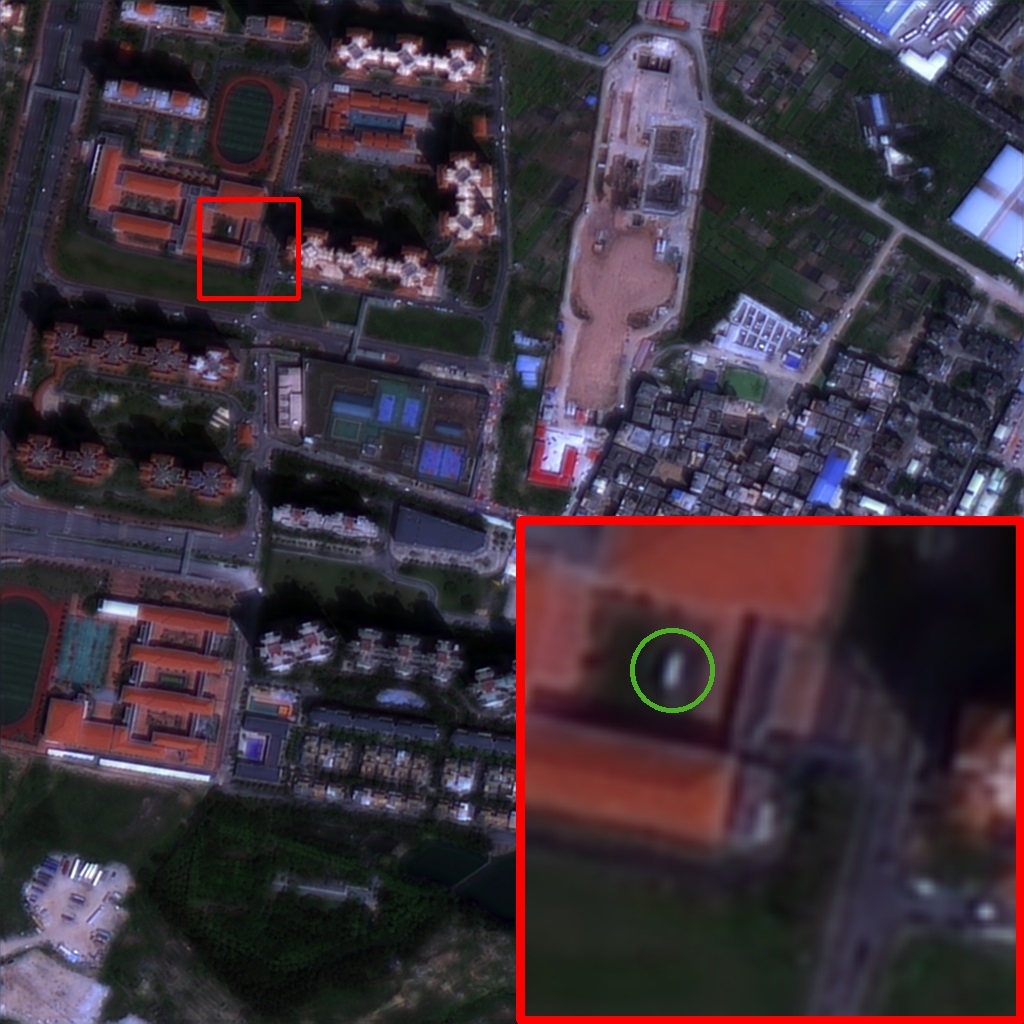}
		\label{fig:visualization_of_original_scale_PNN}
	}
	\subfigure [PanNet]{
		\includegraphics[width=0.3\linewidth]{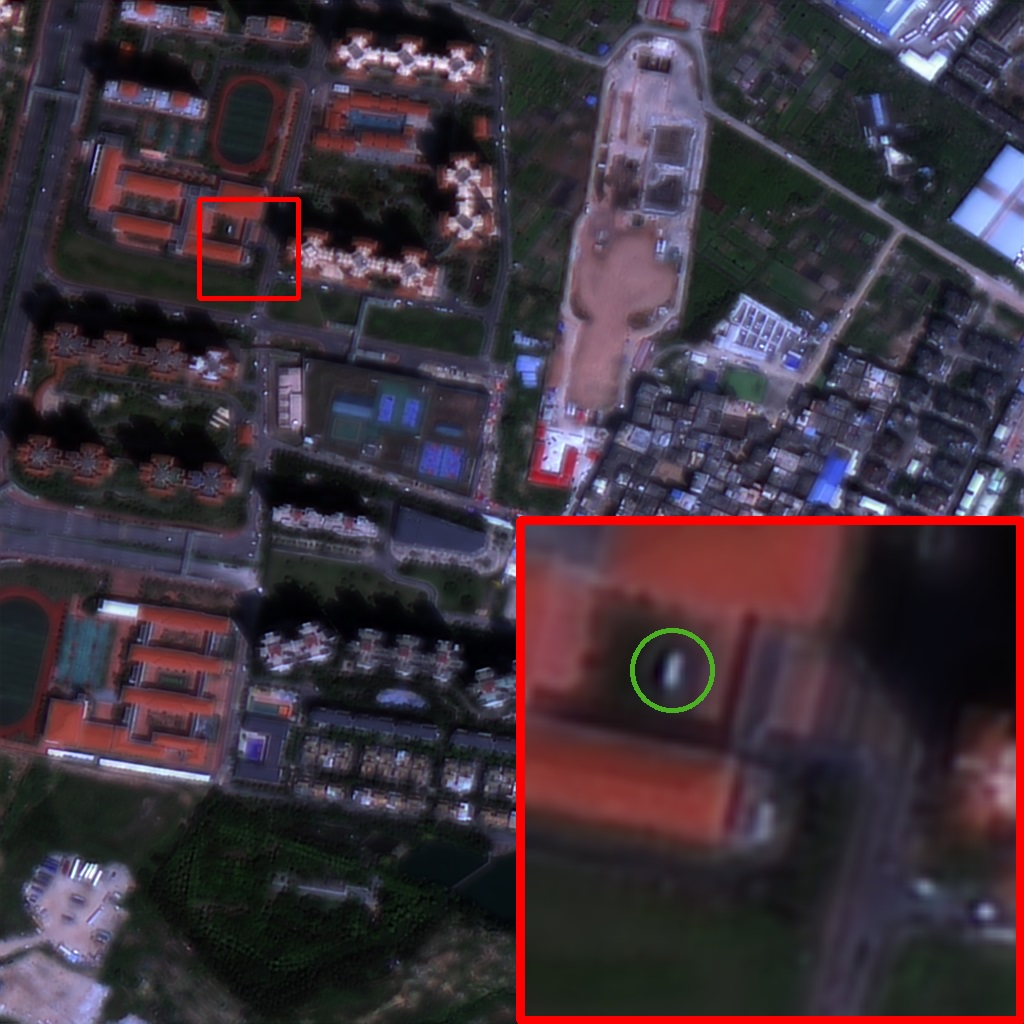}
		\label{fig:visualization_of_original_scale_PanNet}
	}
	\subfigure [PNN+]{
		\includegraphics[width=0.3\linewidth]{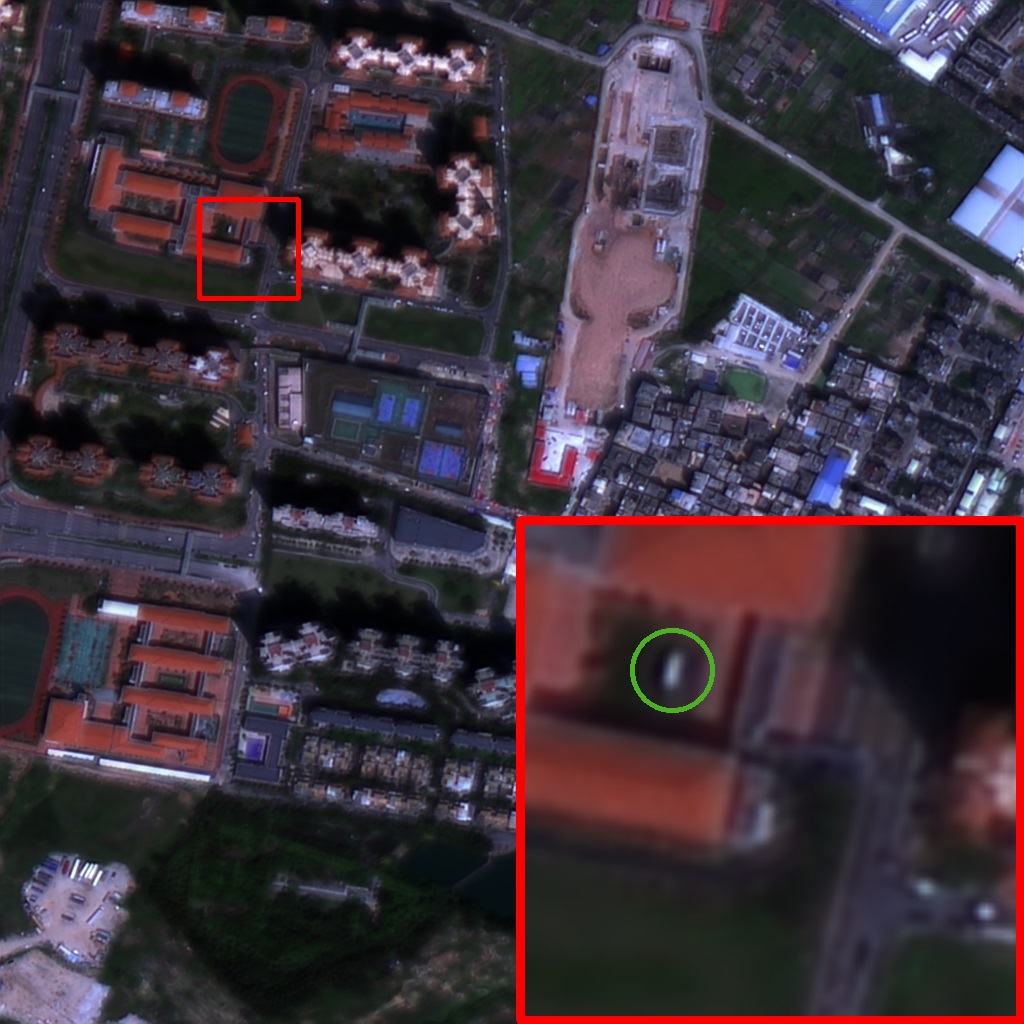}
		\label{fig:visualization_of_original_scale_PNN+}
	}
	\subfigure [DRPNN]{
		\includegraphics[width=0.3\linewidth]{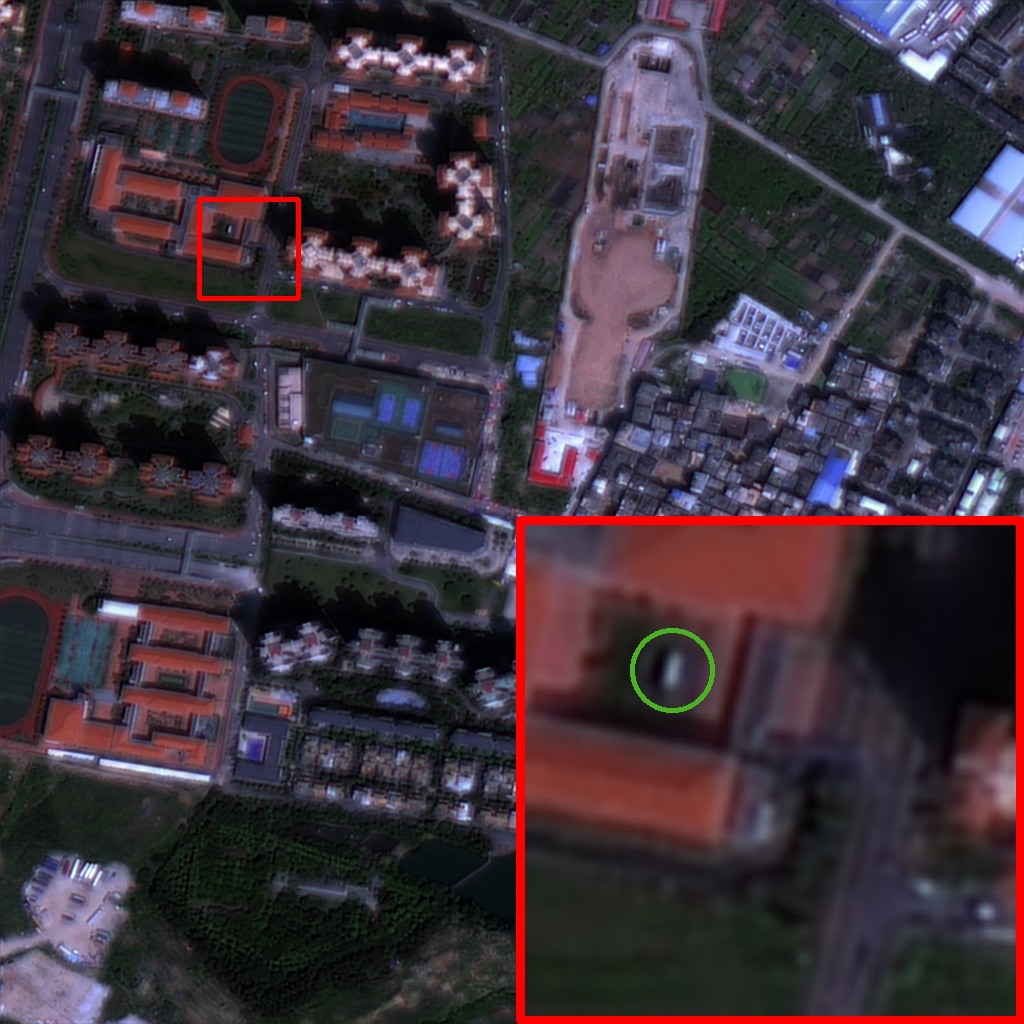}
		\label{fig:visualization_of_original_scale_DRPNN}
	}
	\caption{Pan-sharpened images by different methods on the WorldView-2 dataset.}
	\label{fig:visualization_of_original_scale}
\end{figure}

For \textbf{visual evaluation}, we scale up a small region of all sub-images in Fig. \ref{fig:visualization_of_original_scale}. There is apparent spectral distortion in the green circle in Fig.\ref{fig:visualization_of_original_scale} produced by PNN, PNN+, DRPNN, PanNet, MSDCNN in Figs.\ref{fig:visualization_of_original_scale}(e-i) because these adopt early fusion where they concatenate the PAN image and the bands of MS images as input. The concatenation makes it hard for their model to distinguish each band. The pan-sharpened image Fig. \ref{fig:visualization_of_original_scale_ResTFNet} suffers from spatial distortion as it produced more edges than the PAN image. Only the proposed DCNet can not only make full use of the spatial information provided by the PAN image, but also prevent the distortion of spectral content.

\begin{table}
	\caption{Quantitive results evaluated at the original scale on three datasets. The optimal and the sub-optimal results are in \textcolor{red}{red} and \textcolor{blue}{blue}, respectively.}
	\centering
	\begin{tabular}{ccccc}
		\hline
		Dataset&Method& \makecell[c]{${{D}_{\lambda }}$ } &\makecell[c]{${{D}_{s}}$} & QNR\\
		
		\hline
		 \multirow{7}{*}{IKONOS}&PNN \cite{masi2016pansharpening} &0.0485&0.0608&0.8938\\
		&PNN+ \cite{scarpa2018target}&0.0182&0.0718&0.9114\\
		&DRPNN \cite{wei2017boosting}&0.0219&0.0726&0.9072\\
		&PanNet \cite{yang2017pannet}&0.0235&0.0605&0.9175\\
		&MSDCNN \cite{yuan2018multiscale}&\textcolor{red}{0.0142}&0.0579&\textcolor{blue}{0.9287}\\
		&ResTFNet \cite{liu2020remote}&0.0215&\textcolor{blue}{0.0561}&0.9236\\
		&DCNet(ours)&\textcolor{blue}{0.0174}&\textcolor{red}{0.0521}&\textcolor{red}{0.9314}\\
		\hline
		\multirow{7}{*}{GaoFen-2}&PNN \cite{masi2016pansharpening} &0.0336&0.0592&0.9092\\
		&PNN+ \cite{scarpa2018target}&0.0223&0.0546&0.9244\\
		&DRPNN \cite{wei2017boosting}&0.0657&0.0911&0.8493\\
		&PanNet \cite{yang2017pannet}&0.0394&0.0590&0.9040\\
		&MSDCNN \cite{yuan2018multiscale}&0.0177&0.0421&0.9410\\
		&ResTFNet \cite{liu2020remote}&\textcolor{blue}{0.0141}&\textcolor{blue}{0.0312}&\textcolor{blue}{0.9552}\\
		&DCNet(ours)&\textcolor{red}{0.0128}&\textcolor{red}{0.0307}&\textcolor{red}{0.9569}\\
		\hline
		\multirow{7}{*}{WorldView-2}&PNN \cite{masi2016pansharpening} &0.0277&0.0825&0.8922\\
		&PNN+ \cite{scarpa2018target}&\textcolor{blue}{0.0185}&0.0952&0.8884\\
		&DRPNN \cite{wei2017boosting}&0.0414&0.0904&0.8727\\
		&PanNet \cite{yang2017pannet}&0.0257&0.0980&0.8793\\
		&MSDCNN \cite{yuan2018multiscale}&0.0346&\textcolor{red}{0.0784}&0.8900\\
		&ResTFNet \cite{liu2020remote}&0.0202&\textcolor{blue}{0.0824}&\textcolor{red}{0.8996}\\
		&DCNet(ours)&\textcolor{red}{0.0176}&0.0863&\textcolor{blue}{0.8976}\\
		\hline
		\multicolumn{2}{c}{Ideal value}&0&0&1\\
		\hline
	\end{tabular}
	
	\label{tab:results-original-resolution}
\end{table}

Furthermore, we use the reference-free measurement QNR \cite{alparone2008multispectral} to assess the pan-sharpened images. The QNR index is composed of two components: the spectral distortion index ${{D}_{\lambda }}$ and spatial distortion index ${{D}_{s}}$. Comparison results in Table \ref{tab:results-original-resolution} are obtained by calculating the mean over 50 images for each dataset. The optimal and the sub-optimal results are in red and blue, respectively. It can be concluded that compared with other state-of-the-art methods, the DCNet achieves competitive or superior performance in terms of ${{D}_{\lambda }}$, ${{D}_{s}}$, and QNR metrics. In other words, the DCNet achieves better fusion results in terms of spatial and spectral preservation.

\subsection{Experimental analysis}
\subsubsection{Effect of 2D/3D backbone}

In this part, we will discuss the effectiveness of the dual-channel backbone of 2D/3D architecture for spatial and spectral feature extraction. In contrast to the proposed DCNet that extract spatial and spectral features by a 2D CNN and a 3D CNN, respectively, some existing two-steam methods \cite{shao2018remote} adopt 2D CNNs in both channels. We compare the heterogeneous architecture (denoted as 2D/3D) with the homogeneous counterpart (indicated by 2D/2D) and report the results in Table \ref{tab:backbone}. In the experiment, the 2D/3D backbone has the same convolutional layers as the homogeneous one, and they have the same kernel size at each layer. The S$^2$-CLSTM module is also applied to the compared model to fuse the hierarchical features. The same experimental settings are used for a fair comparison. It can be seen that the network equipped with 2D/3D backbone outperforms the 2D/2D counterpart in terms of all the evaluation indexes, indicating the effectiveness of it.

\begin{table}[htbp]
	\centering
	\caption{Effect of the 2D/3D backbone on the IKONOS, GaoFen-2, and WorldView-2 datasets. The optimal results are in \textcolor{red}{red}.}
	\begin{tabular}{ccccccc}
		\hline
		Dataset & Backbone & Q4/Q8 & UIQI & SAM & ERGAS & SCC \\
		\hline
		\multicolumn{1}{c}{\multirow{2}[2]{*}{IKONOS}} & 2D/2D & 0.6009 & 0.8768 & 4.1365 & 3.7250 & 0.8565 \\
		& 2D/3D & \textcolor[rgb]{ 1,  0,  0}{0.7236} & \textcolor[rgb]{ 1,  0,  0}{0.9242} & \textcolor[rgb]{ 1,  0,  0}{3.6596} & \textcolor[rgb]{ 1,  0,  0}{2.7353} & \textcolor[rgb]{ 1,  0,  0}{0.9502} \\
		\hline
		\multicolumn{1}{c}{\multirow{2}[2]{*}{GaoFen-2}} & 2D/2D & 0.7970 & 0.9517 & 2.9076 & 3.3160 & 0.9160 \\
		& 2D/3D & \textcolor[rgb]{ 1,  0,  0}{0.8707} & \textcolor[rgb]{ 1,  0,  0}{0.9741} & \textcolor[rgb]{ 1,  0,  0}{2.1171} & \textcolor[rgb]{ 1,  0,  0}{2.3805} & \textcolor[rgb]{ 1,  0,  0}{0.9655} \\
		\hline
		\multicolumn{1}{c}{\multirow{2}[2]{*}{WorldView-2}} & 2D/2D & 0.6807 & 0.9207 & 4.8672 & 2.9484 & 0.9420 \\
		& 2D/3D & \textcolor[rgb]{ 1,  0,  0}{0.6982} & \textcolor[rgb]{ 1,  0,  0}{0.9249} & \textcolor[rgb]{ 1,  0,  0}{4.6668} & \textcolor[rgb]{ 1,  0,  0}{2.8387} & \textcolor[rgb]{ 1,  0,  0}{0.9476} \\
		\hline
	\end{tabular}
	\label{tab:backbone}
\end{table}

\subsubsection{Effect of hierarchical feature fusion}

In this experiment, we will test the effectiveness of the hierarchical feature fusion manner. We select different levels for fusion, and the results are listed in Table \ref{tab:hierarchical_feature_fusion}. The level set \{1, 2, 3, 4\} indicates that the DCNet employs the hierarchical fusion manner, in which the S$^2$-CLSTM integrates two channels at each level. \{4\} represents the S$^2$-CLSTM only fuse the spatial and spectral features at single level 4, and \{3,4\}, \{2,3,4\} indicate that the S$^2$-CLSTM merge two channels at two levels 3,4 and at three levels 2,3,4, respectively. For each setting, the same experimental parameters are used, e.g., employing Adam optimizer, the learning rate, the number of epochs, etc. The results show that the model with the hierarchical fusion manner achieves the best results on either spatial indexes or spectral indexes.

\begin{table}[htbp]
	\centering
	\caption{The effect of hierarchical features. The optimal results are in \textcolor{red}{red}.}
	\begin{tabular}{ccccccc}
		\hline
		Dataset & Fusion Levels & Q4/Q8 & UIQI & SAM & ERGAS & SCC \\
		\hline
		\multirow{4}[2]{*}{IKONOS} & \{4\}   & 0.1630 & 0.3103 & 15.4476 & 25.6835 & 0.6087 \\
		& \{3,4\} & 0.3105 & 0.6667 & 10.3325 & 12.7822 & 0.7482 \\
		& \{2,3,4\} & 0.6481 & 0.8989 & 3.9113 & 3.2781 & 0.8997 \\
		& \{1,2,3,4\} & \textcolor[rgb]{ 1,  0,  0}{0.7236} & \textcolor[rgb]{ 1,  0,  0}{0.9242} & \textcolor[rgb]{ 1,  0,  0}{3.6596} & \textcolor[rgb]{ 1,  0,  0}{2.7353} & \textcolor[rgb]{ 1,  0,  0}{0.9502} \\
		\hline
		\multirow{4}[2]{*}{GaoFen-2} & \{4\}   & 0.3085 & 0.3625 & 23.1641 & 30.8456 & 0.6705 \\
		& \{3,4\} & 0.0862 & 0.5095 & 18.5059 & 18.2513 & 0.6677 \\
		& \{2,3,4\} & 0.8181 & 0.9636 & 2.4609 & 2.8779 & 0.9340 \\
		& \{1,2,3,4\} & \textcolor[rgb]{ 1,  0,  0}{0.8707} & \textcolor[rgb]{ 1,  0,  0}{0.9741} & \textcolor[rgb]{ 1,  0,  0}{2.1171} & \textcolor[rgb]{ 1,  0,  0}{2.3805} & \textcolor[rgb]{ 1,  0,  0}{0.9655} \\
		\hline
		\multirow{4}[2]{*}{WorldView-2} & \{4\}   & 0.4827 & 0.1863 & 62.4558 & 32.9168 & 0.7715 \\
		& \{3,4\} & 0.5284 & 0.7866 & 15.4305 & 8.2610 & 0.8565 \\
		& \{2,3,4\} & 0.5536 & 0.8493 & 7.4289 & 6.1733 & 0.8603 \\
		& \{1,2,3,4\} & \textcolor[rgb]{ 1,  0,  0}{0.6982} & \textcolor[rgb]{ 1,  0,  0}{0.9249} & \textcolor[rgb]{ 1,  0,  0}{4.6668} & \textcolor[rgb]{ 1,  0,  0}{2.8387} & \textcolor[rgb]{ 1,  0,  0}{0.9476} \\
		\hline
	\end{tabular}
	\label{tab:hierarchical_feature_fusion}
\end{table}

\subsection{Effect of S$^2$-CLSTM}

To evaluate the effect of the proposed S$^2$-CLSTM fusion module, we replace it with several standard fusion methods, e.g., element-wise summation (sum) fusion, element-wise maximization (max) fusion, element-wise average fusion, element-wise product fusion, and Conv fusion \cite{feichtenhofer2016convolutional}. Similar to the S$^2$-CLSTM, for each of the replaced fusion operations, the fused features at the previous levels, except the last level, are fed back into the two channels, and the fused feature at the last level is directly injected into the reconstruction network. The results are shown in Table \ref{tab:effect_S2-CLSTM}. The results demenstrate that S$^2$-CLSTM fusion strategy can boost the performance of our DCNet on each dataset.

\begin{table}[htbp]
	\centering
	\caption{The results of DCNet with different fusion methods. The optimal results are in \textcolor{red}{red}.}
	\begin{tabular}{ccccccc}
		\hline
		Dataset & Fusion method & Q4/Q8 & UIQI & SAM & ERGAS & SCC \\
		\hline
		\multirow{6}[2]{*}{IKONOS} & Sum   & 0.5649 & 0.8515 & 5.0278 & 4.3132 & 0.8256 \\
		& Max   & 0.5885 & 0.8701 & 4.3609 & 3.8540 & 0.8474 \\
		& Average & 0.6111 & 0.8819 & 4.1098 & 3.6417 & 0.8669 \\
		& Product & 0.5529 & 0.8462 & 5.0558 & 4.2705 & 0.8209 \\
		& Conv  & 0.6402 & 0.8957 & 3.9526 & 3.3367 & 0.8936 \\
		& {S$^2$-CLSTM} & \textcolor[rgb]{ 1,  0,  0}{0.7236} & \textcolor[rgb]{ 1,  0,  0}{0.9242} & \textcolor[rgb]{ 1,  0,  0}{3.6596} & \textcolor[rgb]{ 1,  0,  0}{2.7353} & \textcolor[rgb]{ 1,  0,  0}{0.9503} \\
		\hline
		\multirow{6}[2]{*}{GaoFen-2} & Sum   & 0.7736 & 0.9379 & 3.4308 & 3.8346 & 0.8990 \\
		& Max   & 0.7860 & 0.9441 & 3.2131 & 3.5801 & 0.9072 \\
		& Average & 0.8044 & 0.9565 & 2.7750 & 3.1697 & 0.9231 \\
		& Product & 0.8133 & 0.9611 & 2.5893 & 2.9790 & 0.9312 \\
		& Conv  & 0.8301 & 0.9659 & 2.4278 & 2.7677 & 0.9453 \\
		& {S$^2$-CLSTM} & \textcolor[rgb]{ 1,  0,  0}{0.8707} & \textcolor[rgb]{ 1,  0,  0}{0.9741} & \textcolor[rgb]{ 1,  0,  0}{2.1171} & \textcolor[rgb]{ 1,  0,  0}{2.3805} & \textcolor[rgb]{ 1,  0,  0}{0.9655} \\
		\hline
		\multirow{6}[2]{*}{WorldView-2} & Sum   & 0.6071 & 0.8981 & 5.9973 & 3.646 & 0.9093 \\
		& Max   & 0.6344 & 0.9060 & 5.5230 & 3.3531 & 0.9224 \\
		& Average & 0.6568 & 0.9137 & 5.2045 & 3.1561 & 0.9313 \\
		& Product & 0.6719 & 0.9195 & 4.9014 & 2.9752 & 0.9410 \\
		& Conv  & 0.6771 & 0.9216 & 4.7897 & 2.9107 & 0.9442 \\
		& {S$^2$-CLSTM} & \textcolor[rgb]{ 1,  0,  0}{0.6982} & \textcolor[rgb]{ 1,  0,  0}{0.9249} & \textcolor[rgb]{ 1,  0,  0}{4.6668} & \textcolor[rgb]{ 1,  0,  0}{2.8387} & \textcolor[rgb]{ 1,  0,  0}{0.9476} \\
		\hline
	\end{tabular}%
	\label{tab:effect_S2-CLSTM}%
\end{table}%

\begin{figure}
	\setlength{\abovecaptionskip}{0.cm}
	\centering
	\subfigure [Q]{
		\includegraphics[width=0.46\linewidth]{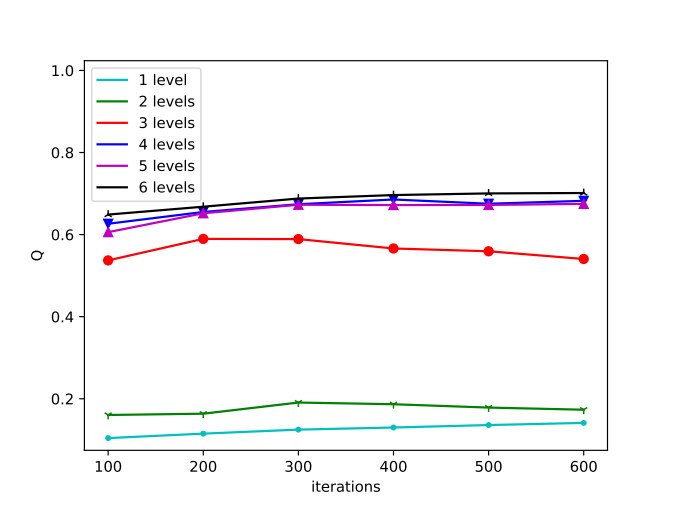}
		\label{fig:levels_Q}
	}
	\subfigure [UIQI]{
		\includegraphics[width=0.46\linewidth]{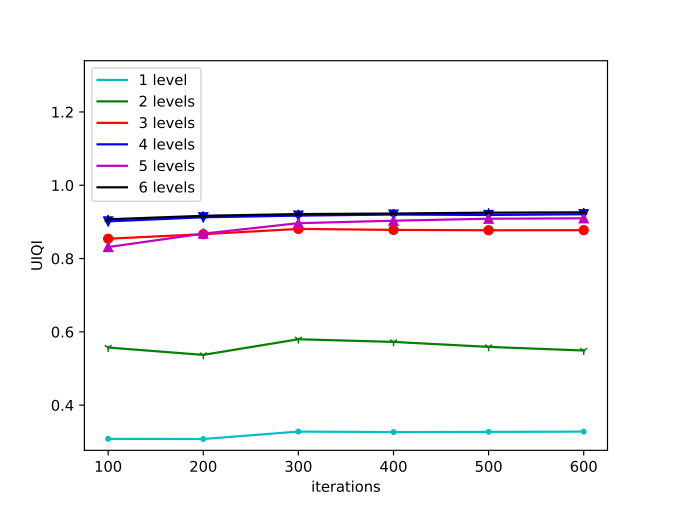}
		\label{fig:levels_UIQI}
	}
	\subfigure [SAM]{
		\includegraphics[width=0.46\linewidth]{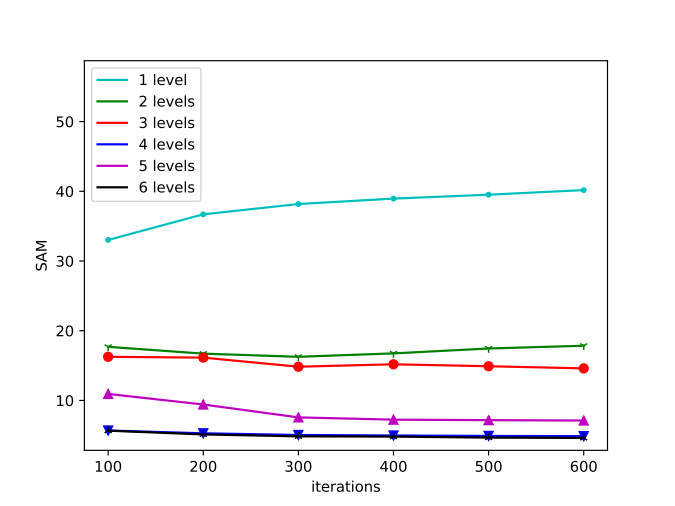}
		\label{fig:levels_SAM}
	}
	\subfigure [ERGAS]{
		\includegraphics[width=0.46\linewidth]{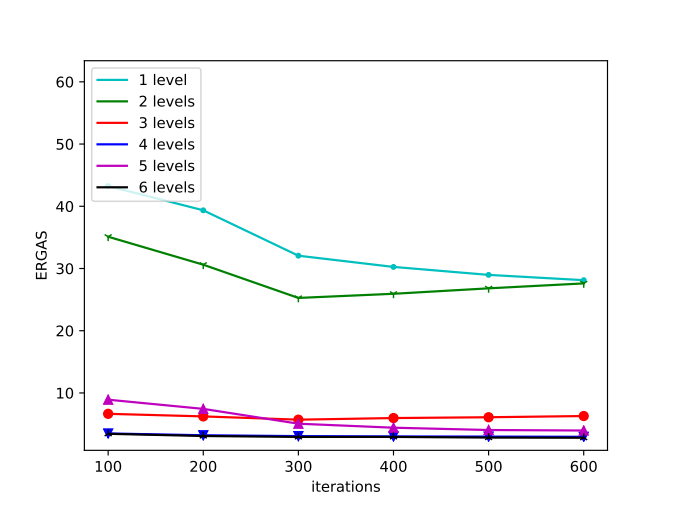}
		\label{fig:levels_ERGAS}
	}
	\subfigure [SCC]{
		\includegraphics[width=0.46\linewidth]{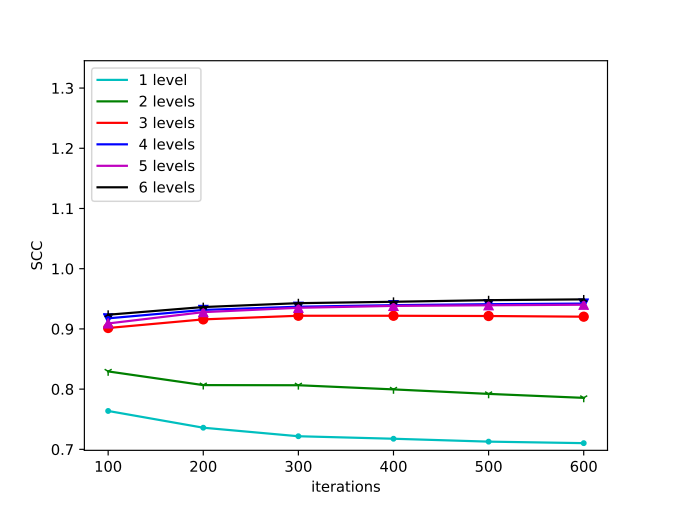}
		\label{fig:levels_SCC}
	}
	\caption{The results of models with different numbers of levels. The models with one, two, three, four, five, and six levels are shown in \textcolor{cyan}{cyan}, \textcolor{green}{green}, \textcolor{red}{red}, \textcolor{blue}{blue}, \textcolor{purple}{purple}, and black}
	\label{fig:effect_of_number_of_levels}
\end{figure}

\subsubsection{Impact of the number of levels}

One of the most critical hyper-parameters in DCNet is the number of levels. It is common knowledge that the nonlinearity of CNNs can be improved by increasing the depth of it, e.g., in \cite{dong2014learning}, deep CNNs have brought prosperous development for the super resolution task. In \cite{wei2017boosting}, the concept of residual learning is introduced to construct a very deep CNN to further improve the performance by making full use of the high nonlinearity of deep CNN models. Thus, the depth of our model needs to be discussed. Experimental results about the impact of the number of levels are shown in Fig. \ref{fig:effect_of_number_of_levels}. The model with 4 levels achieves superior performance when compared to the other four models consisting of 1, 2, 3, 5 levels, respectively, while the model with 6 levels has only small gains. Thus we prefer four levels based model considering the trade-off of the performance and computation cost.

\section{Conclusion}
\label{Conclusion}

In this paper, we have proposed a novel DCNet for pan-sharpening. Instead of employing 2D CNNs for processing both PAN and MS images, we develop a heterogeneous dual-channel backbone with a 2D CNN and a 3D CNN for spatial and spectral information extraction, respectively. The 3D CNN in spectral channel avoid the mixing of different spectrums, which facilitates the HR-MS image reconstruction. The S$^2$-CLSTM fuses spatial and spectral information at each level via bi-directional lateral connections and expoit the multi-level correlation via the cell state. Compared with state-of-the-art methods, the proposed network demonstrates the potent ability for feature extraction and fusion of spatial and spectral information on three datasets. For future work, within the current framework, loss functions and unsupervised methods will be explored for the real-world task.

\section*{Declaration of Competing Interest}
The authors declare that they have no known competing financial interests or personal relationships that could have appeared to influence the work reported in this paper.

\section*{CRediT authorship contribution statement}
Dong Wang: Conceptualization, Methodology, Data curation, Software, Writing - Original draft preparation. Yunpeng Bai: Visualization, Investigation, Validation, Writing- Reviewing and Editing. Ying Li: Supervision, Software, Writing- Reviewing and Editing.

\section*{Acknowledgements}
The work was supported in part by the National Natural Science Foundation of China(61871460), the Shaanxi Provincial Key R\&D Program(2020KW-003), and the Fundamental Research Funds for the Central Universities (3102019ghxm016).



  \bibliographystyle{elsarticle-num} 
  \biboptions{square,numbers,sort&compress}
  \bibliography{CLSTM}


%%%% \bibitem{label}
%%%% Text of bibliographic item
%%
%%\bibitem{}
%%
%%\end{thebibliography}
\end{document}